\RequirePackage{fix-cm}
\documentclass[twocolumn,epjc3]{svjour3}  
\smartqed  
\RequirePackage{graphicx}
\journalname{Eur. Phys. J. C}
\usepackage{amssymb}
\newcommand{\be}{\begin{eqnarray}}
\newcommand{\ee}{\end{eqnarray}}

\usepackage{color}

\begin{document}

\title{Longitudinal momentum densities in transverse plane for nucleons}
\author{Chandan Mondal}
\institute{Department of Physics, Indian Institute of Technology Kanpur, Kanpur-208016, India.}
\date{Received: date / Revised version: date}
%
\maketitle

\begin{abstract}
We present a study of longitudinal momentum densities ($p^+ \rm densities$) in the transverse impact parameter space for $u$ and $d$ quarks in both unpolarized and transversely polarized nucleons by taking a two dimensional Fourier transform of the gravitational form factors with respect to the momentum transfer in the transverse direction. The gravitational form factors are obtained by the second moments of GPDs. Here we consider the GPDs of two different soft-wall models in AdS/QCD correspondence.

\end{abstract}


\vskip0.2in
\noindent
\section{ Introduction}
Recently, AdS/QCD has emerged as one of the most encouraging techniques to unravel the structure of hadrons.  The AdS/CFT duality\cite{maldacena} relates a gravity theory in $AdS_{d+1}$ to a conformal theory   at the  $d$ dimensional boundary.  There are many applications of AdS/CFT duality to investigate the QCD phenomena\cite{PS,costa1}. To compare with the QCD, we needs to break the conformal invariance.  An IR cutoff is set at $z_0=1/\Lambda_{QCD}$ in  the hard-wall model while in soft-wall model, a confining potential in $z$ is introduced which breaks the conformal invariance and allows QCD mass scale and confinement. 
There is an exact correspondence between the holographic variable $z$ and the light-cone transverse variable $\zeta$ which measures the separation of the quark and  gluonic constituents in the hadron\cite{BT00,BT01}.
 The AdS/QCD for the baryon  has been developed by  several groups \cite{BT00,BT01,katz,SS,AC,ads1,ModelII,ads2}.
Though this correspondence gives only the semi-classical approximation of QCD, so far this method has  been successfully applied to  describe many hadron properties e.g., hadron mass spectrum, parton distribution functions, GPDs, meson and nucleon form factors, charge densities, structure functions etc\cite{AC,ads1,ModelII,ads2,AC4,BT1,BT1b,BT2,vega,CM,CM2,CM3,HSS}. The first application in AdS/QCD to nucleon resonances have been reported in \cite{reso}. AdS/QCD wave functions are used to predict the experimental data of $\rho$ meson electroproduction \cite{forshaw}. 
AdS/QCD correspondence has also been successfully applied in the meson sector to predict isospin asymmetry and branching ratio for the $B\to K^*\gamma$ decays \cite{ahmady2}, the branching ratio for decays of $\bar{B^0}$ and $\bar{B_s^0}$ into $\rho$ mesons \cite{ahmady1}, transition form factors\cite{AC2,ahmady3}, etc. There are many other applications in the baryon sector e.g.,  semi-empirical hadronic momentum density distributions in the transverse plane have been calculated in\cite{abidin08}, in \cite{hong2}, the form factor of spin $3/2$ baryons ($\Delta$ resonance) and the transition form factor between $\Delta$ and nucleon have been reported, an AdS/QCD model has been proposed to study the baryon spectrum at finite temperature\cite{li} etc. Recently, it has been shown that there exit a precise mapping between the superconformal quantum mechanics and AdS/QCD \cite{BT_new1}. The superconformal quantum mechanics together with light-front AdS/QCD, has resolved the importance of conformal symmetry and its breaking within the algebraic structure for understanding the confinement mechanism of QCD \cite{BT_new2,BT_new3}.


Matrix elements of the energy momentum tensor ($T^{\mu \nu}$) relate the gravitational form factors(GFFs) which play an important role in hadron physics. For spin $1/2$ particles, similar to the electromagnetic form factors, the GFFs $A(Q^2)$ and $B(Q^2)$ can be obtained from the helicity conserved and helicity-flip matrix elements of the $T^{++}$ tensor current. $A(Q^2)$ and $B(Q^2)$ are analogous to $F_1(Q^2)$ (Dirac) and $F_2(Q^2)$ (Pauli) form factors for the $J^+$ vector current. The helicity conserved GFF $A(Q^2)$ allows us to measure the momentum fractions carried by each constituent of a hadron. 
Ji's sum rule states that $2\langle J_q\rangle=A_q(0)+B_q(0)$ \cite{ji97}. Thus, one has to measure the 
GFFs $A(Q^2)$ and $B(Q^2)$ to find the quark contributions to the nucleon spin. 
In \cite{BTgrav}, Brodsky and Ter\'{a}mond have established the existence of the correspondence between the matrix elements of the energy-momentum tensor of the fundamental hadronic constituents in QCD with the transition amplitudes describing the interaction of string modes in AdS space with an external graviton field which propagates in the AdS interior. They have shown that the GFFs, calculated in light-front as well as in AdS space using two parton hadronic state are equivalent. The GFF $A(Q^2)$ for nucleon in AdS/QCD considering both the hard-wall where the AdS geometry is cut off at $z_0=1/\Lambda_{QCD}$ and the soft-wall model where the geometry is smoothly cut off by a background dilaton field has been evaluated in \cite{AC}. The GFFs of vector mesons in a holographic model of QCD have been studied in \cite{AC4} whereas the GFFs for pion and axial-vector mesons sector in the AdS/QCD hard-wall model have been reported in \cite{AC5}. In both cases, the authors have reported the sum rules connecting the 
GFFs to the corresponding GPDs. 

The charge and magnetization densities inside a nucleon are related to Fourier transforms of charge(Dirac) and magnetic(Pauli) FFs. In a similar fashion, one can map the distribution of longitudinal momentum density within a hadron to Fourier transforms of the GFFs. The momentum density distributions within nucleons and similar distributions for spin-1 objects based on theoretical results from the AdS/QCD correspondence have been calculated in \cite{abidin08}. For nucleons momentum densities, the authors of the Ref.\cite{abidin08} have evaluated the GFFs by the second moment of the GPDs for the ``modified Regge model'' with quarks and gluon distributions of MRST2002 \cite{MRST2002}. A nice comparative study of charge and momentum density distributions have been done in \cite{selyugin} where the authors have used a different $t$-dependence of GPDs from \cite{abidin08} with the same quarks distributions of MRST2002. They have calculated the GFFs and momentum densities of nucleon considering only the valence 
quarks contributions. Recently, a 
transverse spin sum rule \cite{ji12,ji3,JXY,Leader1,HKMR,hari} connecting the relevant GFFs $A(Q^2)$, $B(Q^2)$ and $\bar{C}(Q^2)$ has been verified using a light
front quark-diquark model in AdS/QCD \cite{CMA}. The longitudinal momentum densities have also been evaluated for both the unpolarized and the transversely polarized nucleons in this article~\cite{CMA}. 
 
There are two different holographic QCD  models for nucleon FFs developed by Abidin and Carlson\cite{AC} and Brodsky and Teramond\cite{BT2}. A detailed analysis of the transverse charge and anomalous magnetization densities in both these holographic models have been presented in \cite{CM3}. It is interesting and instructive to study the flavor GFFs as well as the flavor structures of nucleons momentum densities in transverse plane in holographic QCD. In this work, we present a comparative study of the flavor GFFs in both the models. We compare the AdS/QCD results of GFFs with the           
results of a phenomenological model \cite{selyugin}. We evaluate the flavor longitudinal momentum density distributions in transverse plane for both unpolarized and transversely polarized nucleons in both the models. 
 
The paper is organized as follows. A brief description of the two soft-wall AdS/QCD models has been given in Sec.\ref{GFFs}. We also present the flavor GFFs in this section. In Sec.\ref{long_mom_den}, the flavor longitudinal momentum densities for both unpolarized and transversely polarized nucleon have been discussed. Then we provide a brief summary in Sec.\ref{summary}. The longitudinal momentum density for nucleon in a soft-wall as well as in a hard-wall AdS/QCD models has been evaluated in the 
appendix.
\section{Gravitational form factors}\label{GFFs}
GFFs can be obtained by the $x$ moments of the GPDs. In this section we briefly review the prescription to extract GPDs from the nucleon Dirac and Pauli form factors(FFs) in the two different AdS/QCD soft-wall models of nucleon electromagnetic form factors proposed by Brodsky and Ter\'{a}mond~\cite{BT2} and Abidin and Carlson~\cite{AC}. 
\subsection{Model I}
Model-I refers to the soft-wall model of AdS/QCD developed by Brodsky and Ter\'{a}mond for the nucleon form factors \cite{BT2} and the GDPs evaluated in \cite{CM}. 
The relevant AdS/QCD action for the fermion field is written as 
\be
S&=&\int d^4x dz \sqrt{g}\Big( \frac{i}{2}\bar\Psi e^M_A\Gamma^AD_M\Psi -\frac{i}{2}(D_M\bar{\Psi})e^M_A\Gamma^A\Psi\nonumber\\
&&-\mu\bar{\Psi}\Psi-V(z)\bar{\Psi}\Psi\Big),\label{action}
\ee
where $e^M_A=(z/R)\delta^M_A$ is the inverse  vielbein and $V(z)$ is the confining potential which breaks the conformal invariance and 
 $R$ is the AdS radius. One can derive the Dirac equation in AdS from the above action as
\be
i\Big(z \eta^{MN}\Gamma_M\partial_N+\frac{d}{2}\Gamma_z\Big)\Psi -\mu R\Psi-RV(z)\Psi=0.\label{ads_DE}
\ee
In $d=4$ dimensions, $\Gamma_A=\{\gamma_\mu, -i\gamma_5\}$.
To map with the light front wave equation, one identifies $z\to\zeta$ (light front transverse impact variable) and substitutes $\Psi(x,\zeta)=e^{-iP\cdot x}\zeta^2\psi(\zeta)u(P)$ in Eq.(\ref{ads_DE}) and sets $\mid \mu R\mid=\nu+1/2$ where  $\nu$ is related with the orbital angular momentum by $\nu=L+1$ .
For linear confining potential  $U(\zeta)=(R/\zeta)V(\zeta)=\kappa^2\zeta$, one gets the light front wave equation for the baryon 
in $2\times 2$ spinor representation as
\be
\big(-\frac{d^2}{d\zeta^2}-\frac{1-4\nu^2}{4\zeta^2}+\kappa^4\zeta^2&+&2(\nu+1)\kappa^2\Big)\psi_+(\zeta)\nonumber\\&=&{\cal{M}}^2\psi_+(\zeta),\\
 \big(-\frac{d^2}{d\zeta^2}-\frac{1-4(\nu+1)^2}{4\zeta^2}&+&\kappa^4\zeta^2+2\nu\kappa^2\Big)\psi_-(\zeta)\nonumber\\&=&{\cal{M}}^2\psi_-(\zeta),
 \ee 
 which leads to the AdS solutions of nucleon wave-functions $\psi_+(z)$ and $\psi_-(z)$ corresponding to different orbital angular momentum $L^z=0$ and $L^z=+1$ \cite{BT2}
\be
\psi_+(\zeta)\sim\psi_+(z) &=& \frac{\sqrt{2}\kappa^2}{R^2}z^{7/2} e^{-\kappa^2 z^2/2}\label{psi+},\\
\psi_-(\zeta)\sim\psi_-(z) &=& \frac{\kappa^3}{R^2}z^{9/2} e^{-\kappa^2 z^2/2}\label{psi-}.
\ee 
The Dirac form factors in this model are obtained by the SU(6) spin-flavor symmetry and given by 
\be
F_1^p(Q^2)&=&R^4\int \frac{dz}{z^4} V(Q^2,z)\psi^2_+(z),\label{F1p}\\
F_1^n(Q^2) &=& -\frac{1}{3}R^4\int \frac{dz}{z^4} V(Q^2,z)(\psi^2_+(z)-\psi^2_-(z)).\label{F1n}
\ee
The Pauli form factors for the nucleons are modeled in this model as   
\be
F_2^{p/n}(Q^2) =  \kappa_{p/n}R^4 \int \frac{dz}{z^3}\psi_+(z) V(Q^2,z) \psi_-(z).\label{F2}
\ee
The Pauli form factors are normalized to $F_2^{p/n}(0) = \kappa_{p/n}$ where $\kappa_{p/n}$ are the anomalous magnetic moment of proton/neutron. It should be noted that the Pauli form factor is not mapped properly in this model.  In the light front quark model, Pauli form factor is defined as the spin flip matrix element of $J^+$ current but the AdS action in Eq.(\ref{action}) is unable to produce this form factor and it is put in for phenomenological purposes. The bulk-to-boundary propagator for soft wall model
is given by \cite{Rad,BT2}
\be
\! V(Q^2,z)=\kappa^2z^2\int_0^1\!\frac{dx}{(1-x)^2} x^{Q^2/(4\kappa^2)} e^{-\kappa^2 z^2 x/(1-x)}.\label{V}
\ee
Here we use the value $\kappa=0.4 ~\rm GeV$ which is fixed by fitting the ratios of Pauli and Dirac form factors for proton with the experimental data \cite{CM,CM2}. We refer the formulas for the form factors given in Eqs.(\ref{F1p},\ref{F1n} and \ref{F2}) as Model I.

\begin{figure*}[htbp]
\begin{minipage}[c]{0.98\textwidth}
\small{(a)}
\includegraphics[width=7cm,height=5.5cm,clip]{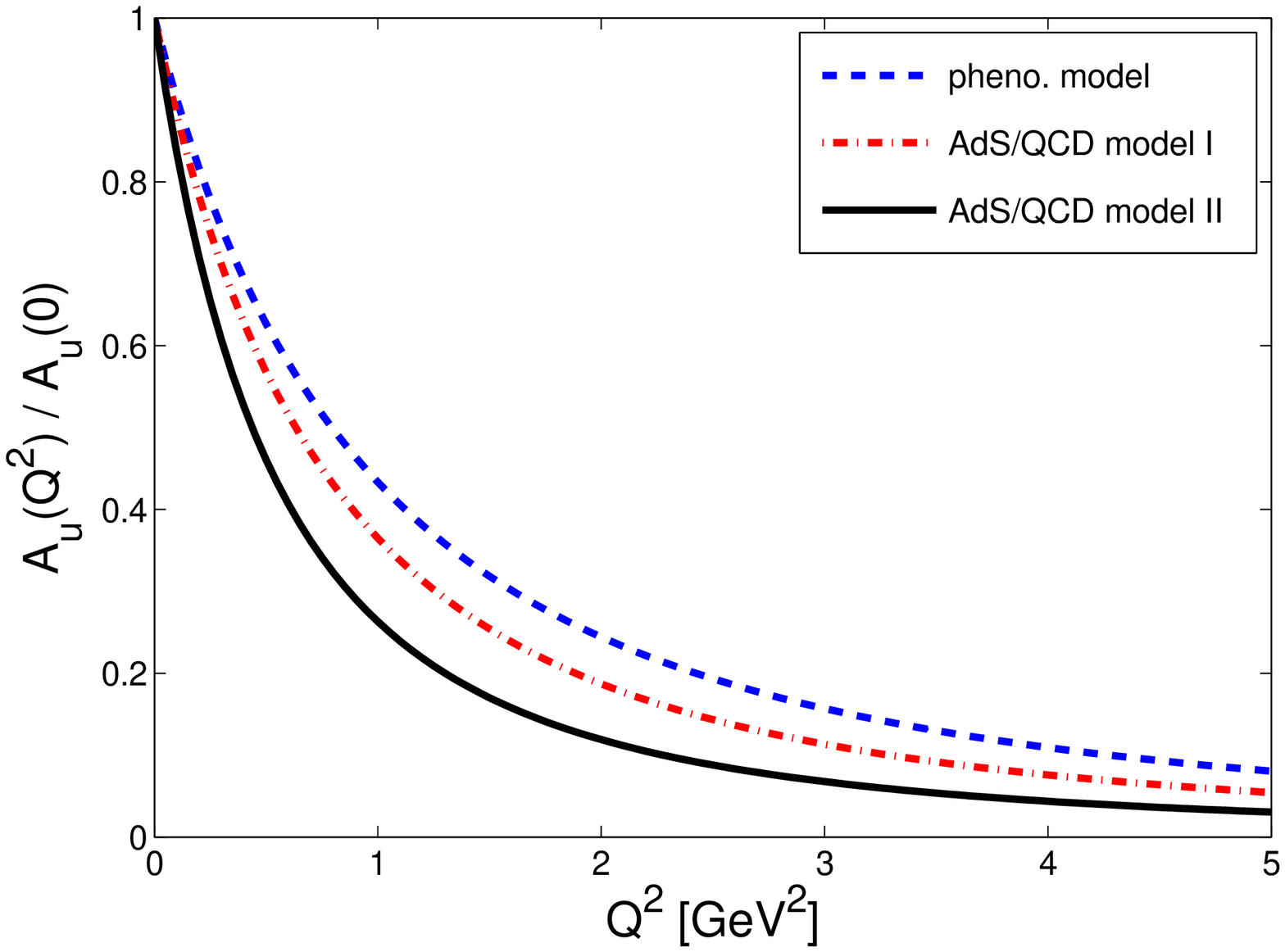}
\hspace{0.1cm}%
\small{(b)}\includegraphics[width=7cm,height=5.5cm,clip]{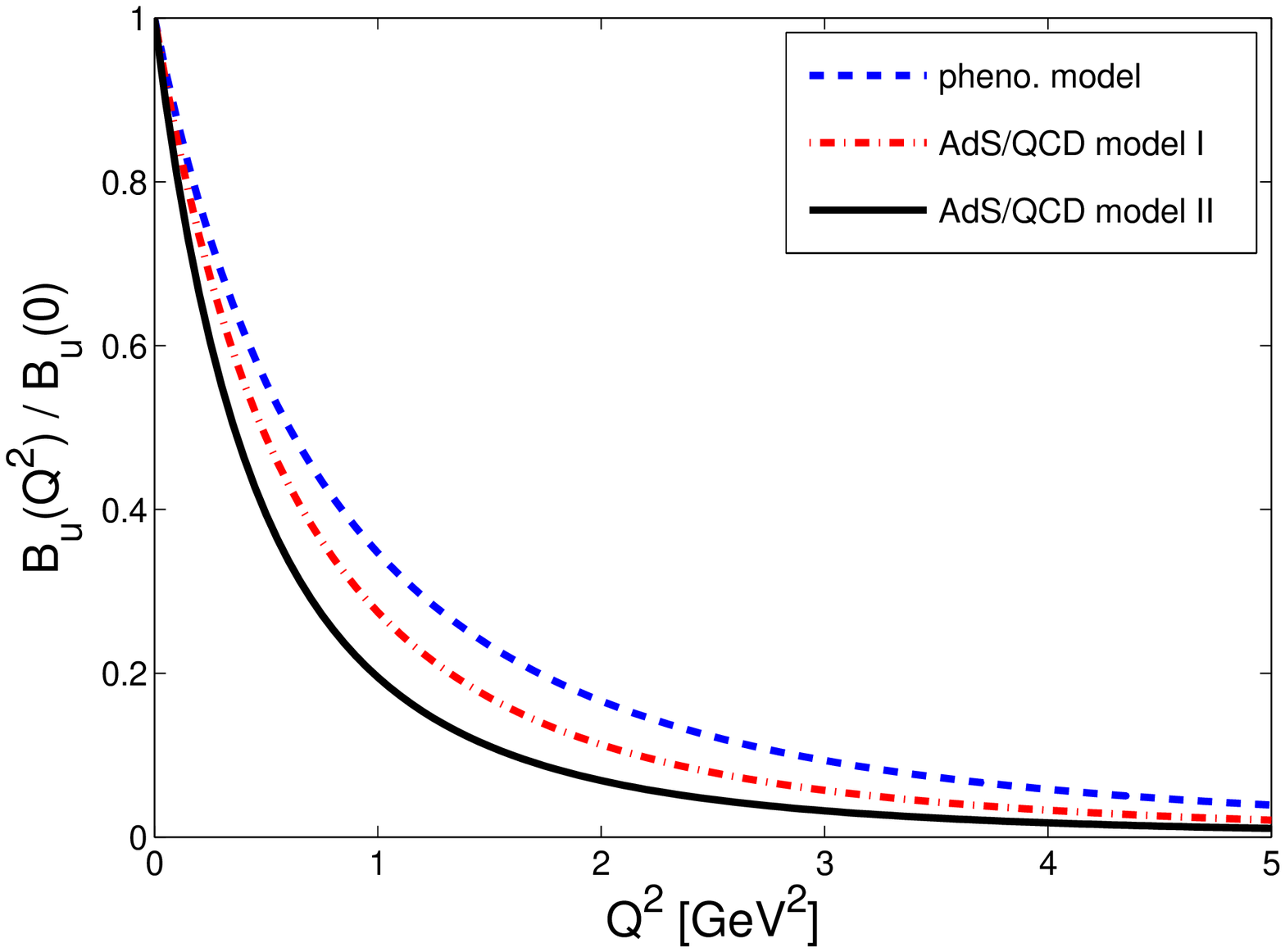}
\end{minipage}
\begin{minipage}[c]{0.98\textwidth}
\small{(c)}
\includegraphics[width=7cm,height=5.5cm,clip]{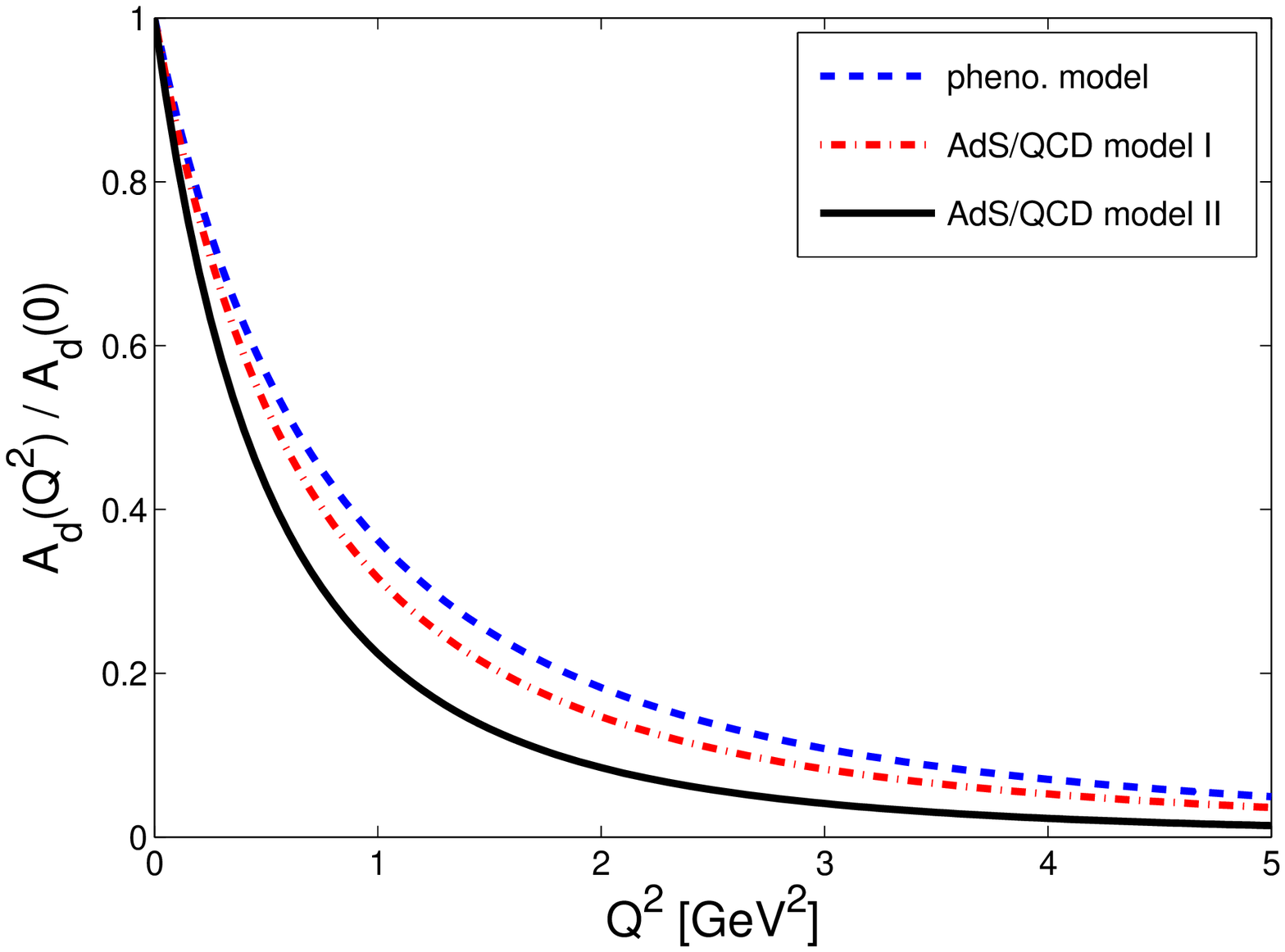}
\hspace{0.1cm}%
\small{(d)}\includegraphics[width=7cm,height=5.5cm,clip]{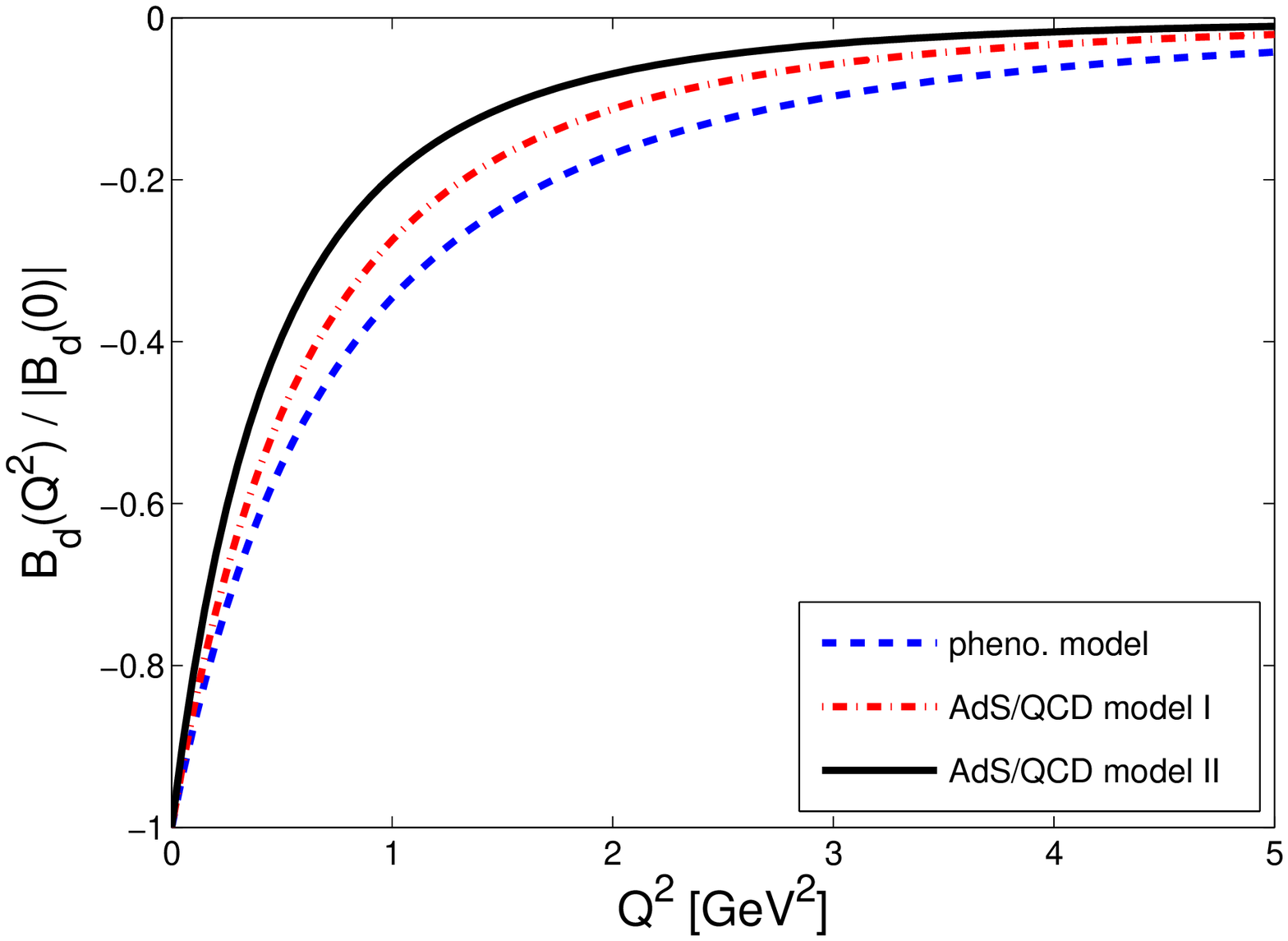}
\end{minipage}
\caption{\label{gffs}(Color online) Plots of gravitational form factors $A(Q^2)$ and $B(Q^2)$ for $u$ and $d$ quarks. The form factors are normalized to unity, upper panel for $u$ quark and lower panel for $d$ quark. Blue dashed line represents the result of a phenomenological model~\cite{selyugin}.}
\end{figure*}
\begin{figure*}[htbp]
\begin{minipage}[c]{0.98\textwidth}
\small{(a)}
\includegraphics[width=7cm,height=5.5cm,clip]{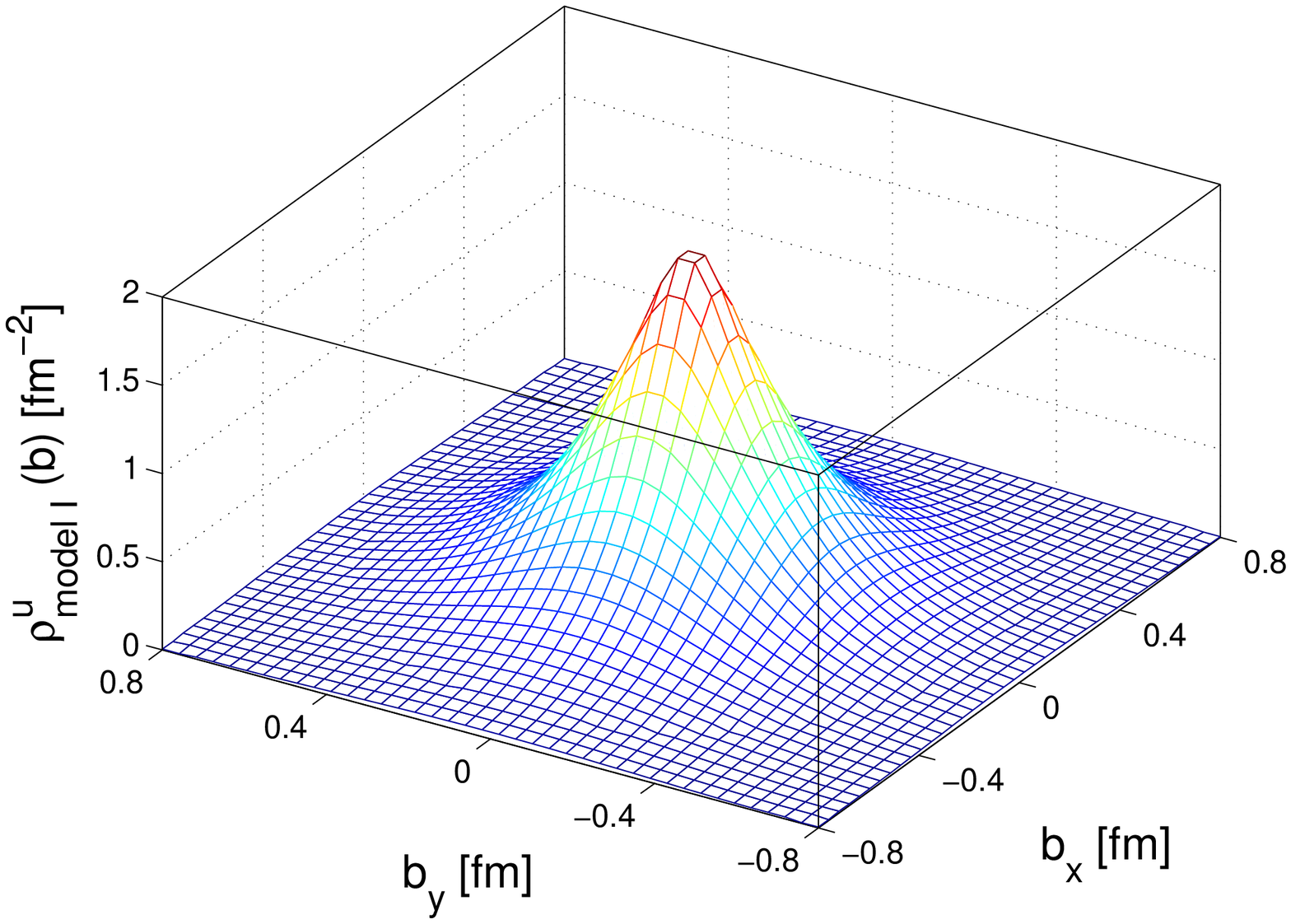}
\hspace{0.1cm}%
\small{(b)}\includegraphics[width=7cm,height=5.5cm,clip]{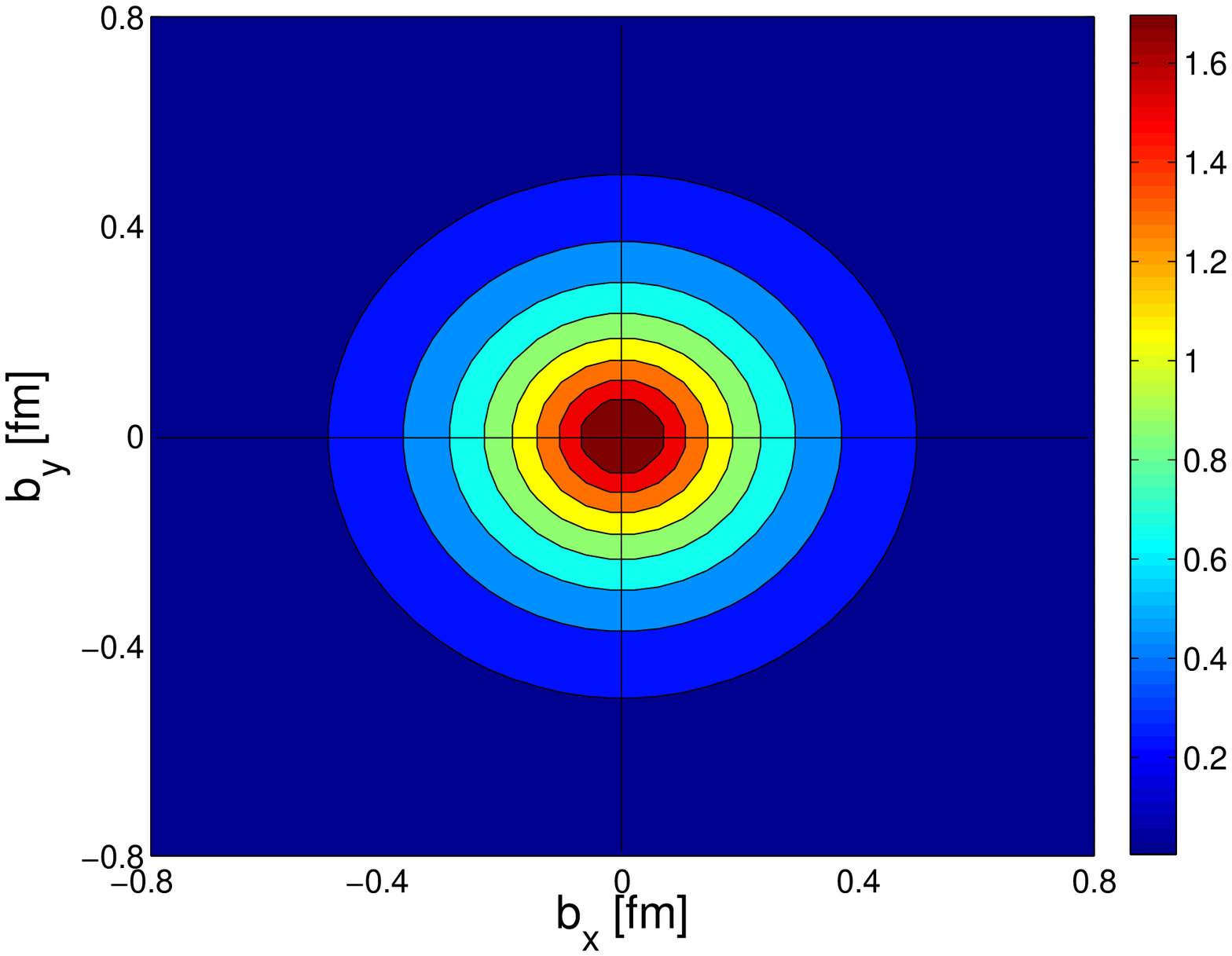}
\end{minipage}
\begin{minipage}[c]{0.98\textwidth}
\small{(c)}
\includegraphics[width=7cm,height=5.5cm,clip]{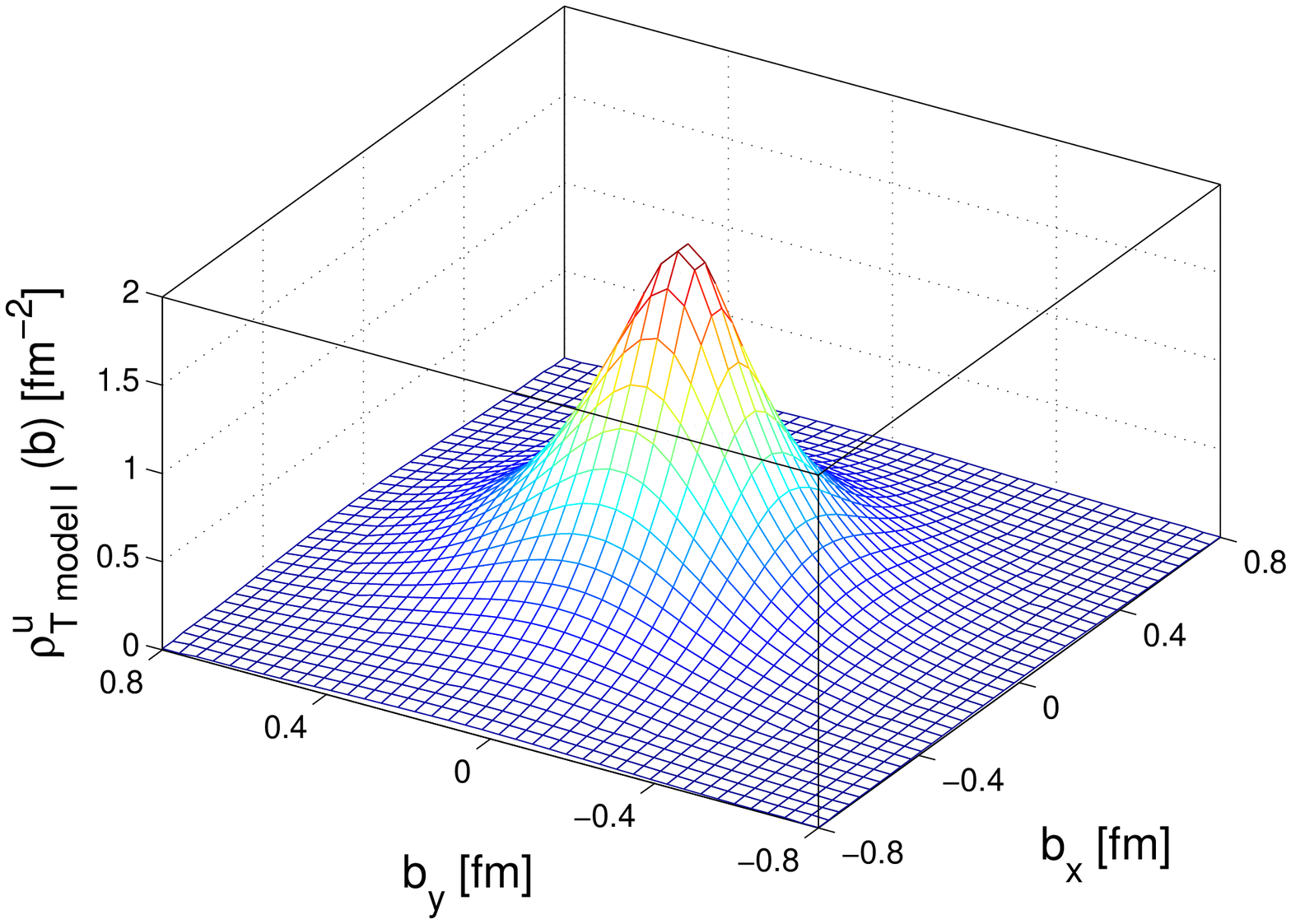}
\hspace{0.1cm}%
\small{(d)}\includegraphics[width=7cm,height=5.5cm,clip]{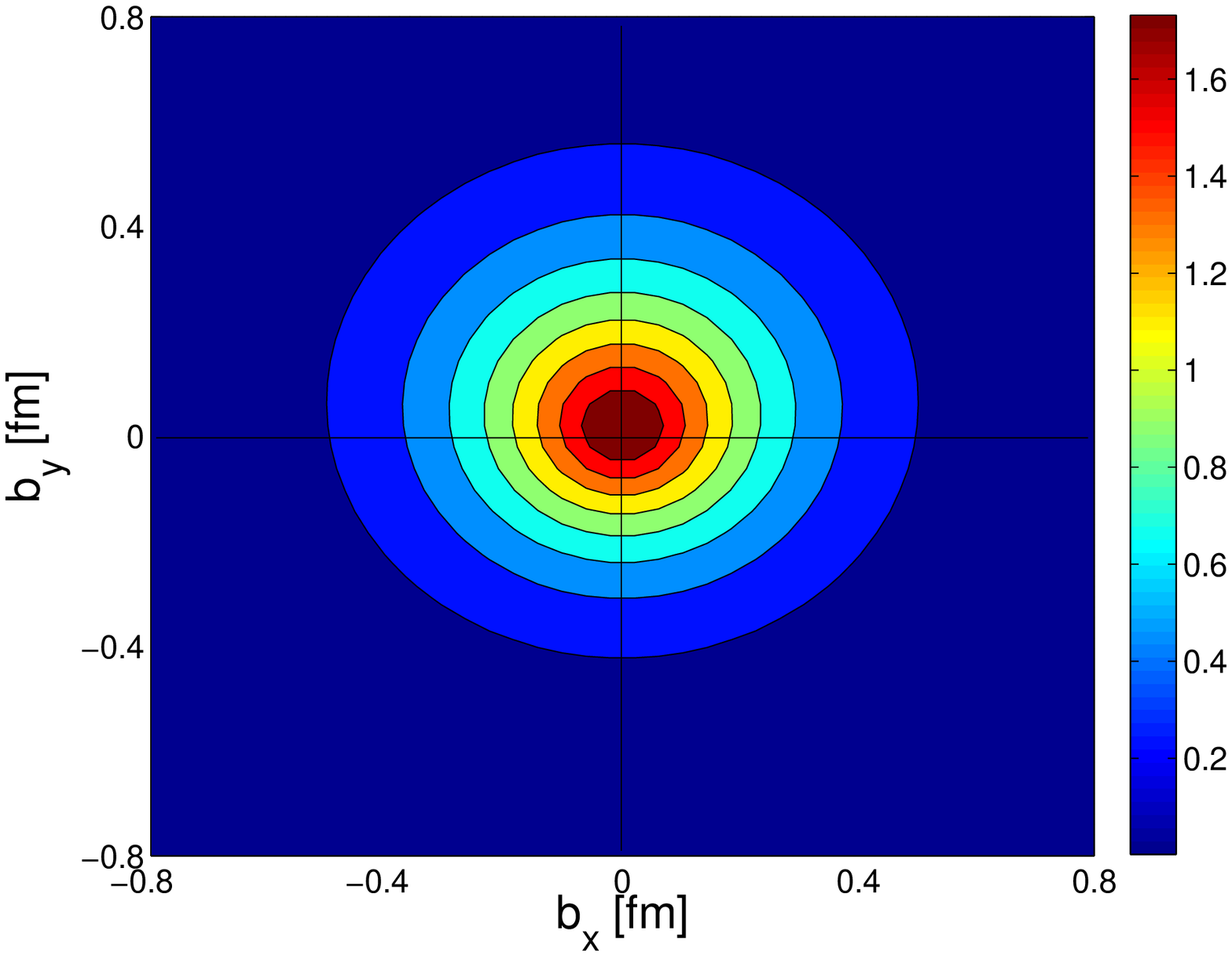}
\end{minipage}
\caption{\label{den_uI}(Color online) The longitudinal momentum densities for $u$ quark in the transverse plane for AdS/QCD Model I, upper panel  for unpolarized proton, lower panel for proton polarized along $x$-direction. (b) and (d) are the top view of (a) and (c) respectively.}
\end{figure*} 
\begin{figure*}[htbp]
\begin{minipage}[c]{0.98\textwidth}
\small{(a)}
\includegraphics[width=7cm,height=5.5cm,clip]{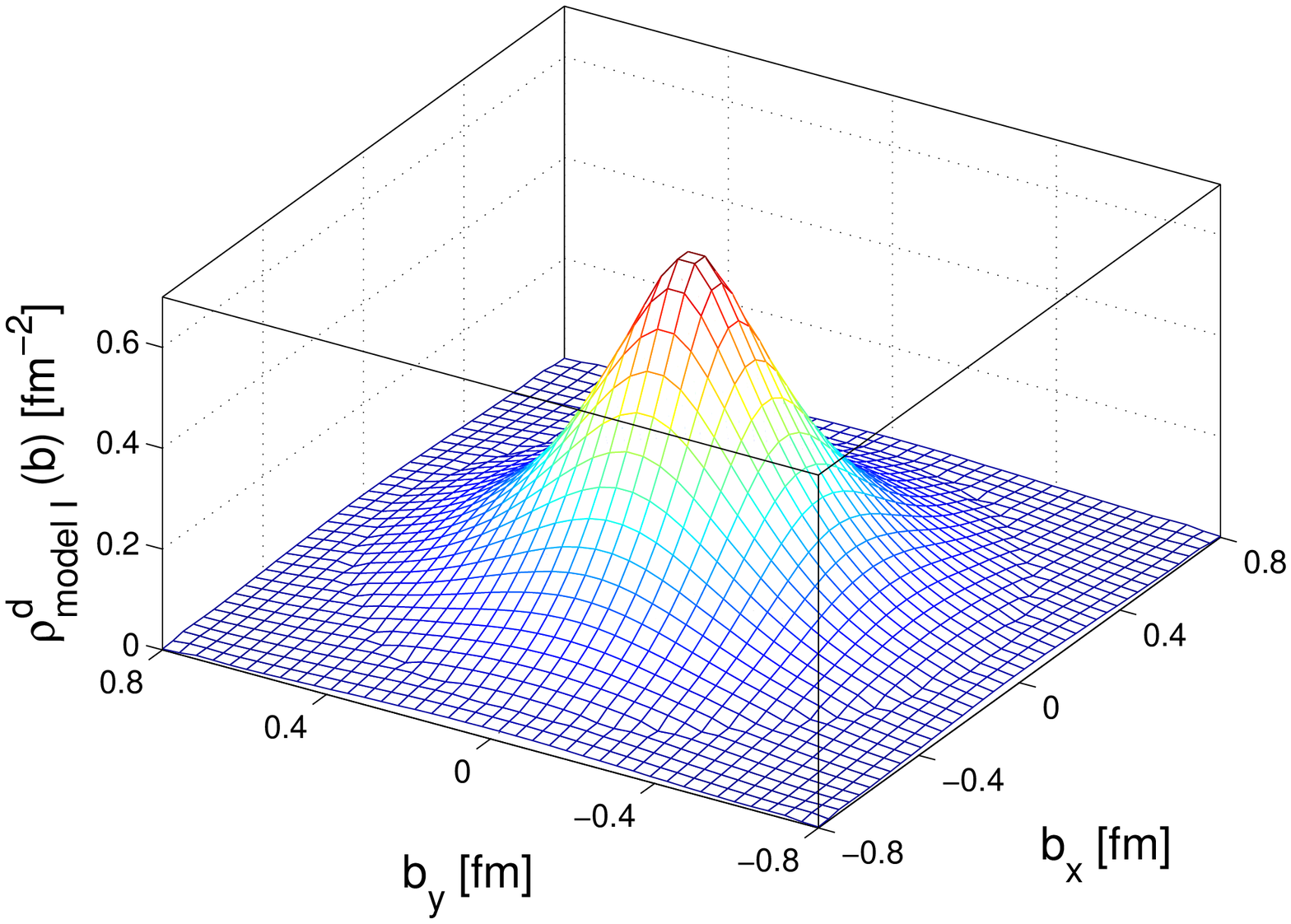}
\hspace{0.1cm}%
\small{(b)}\includegraphics[width=7cm,height=5.5cm,clip]{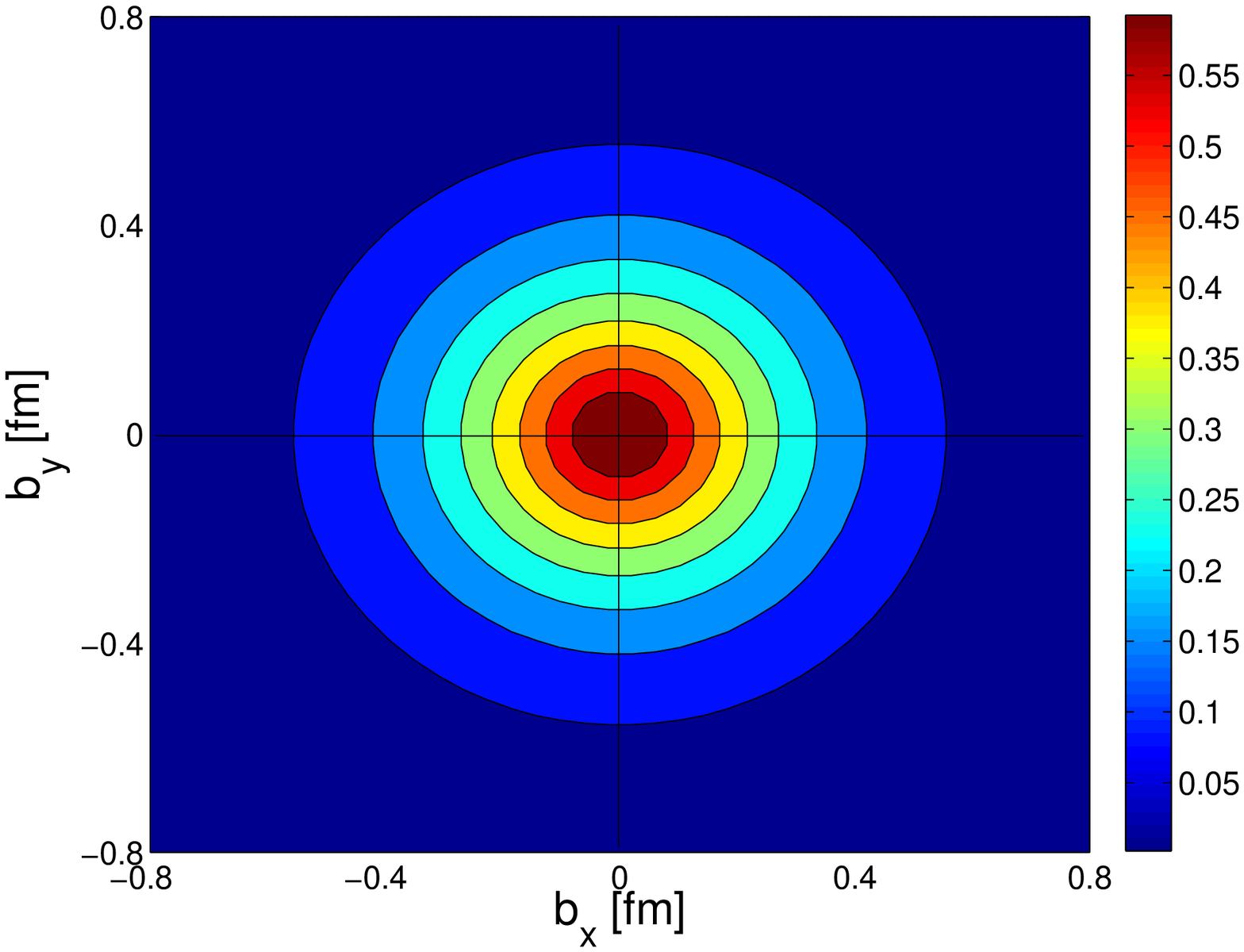}
\end{minipage}
\begin{minipage}[c]{0.98\textwidth}
\small{(c)}
\includegraphics[width=7cm,height=5.5cm,clip]{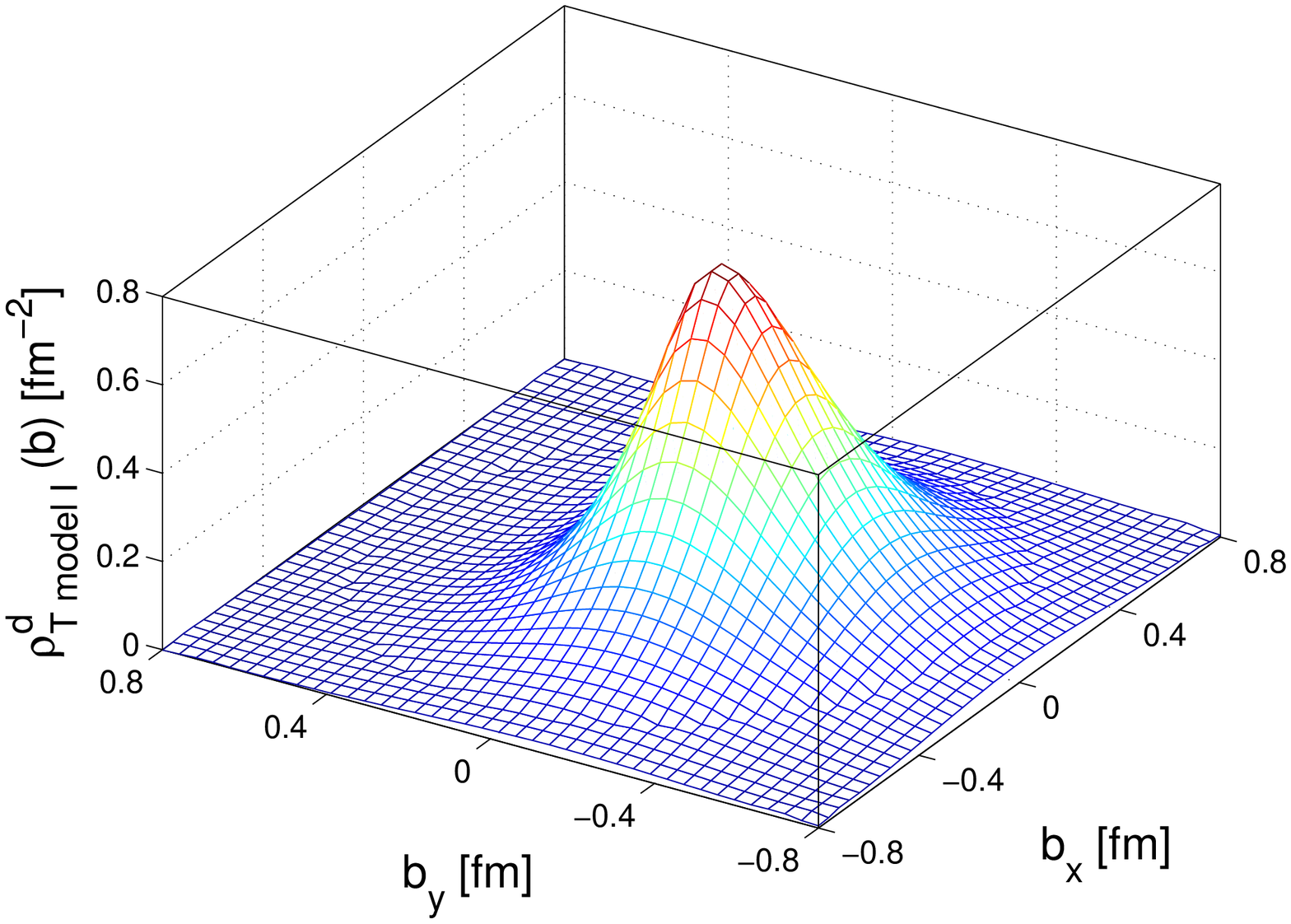}
\hspace{0.1cm}%
\small{(d)}\includegraphics[width=7cm,height=5.5cm,clip]{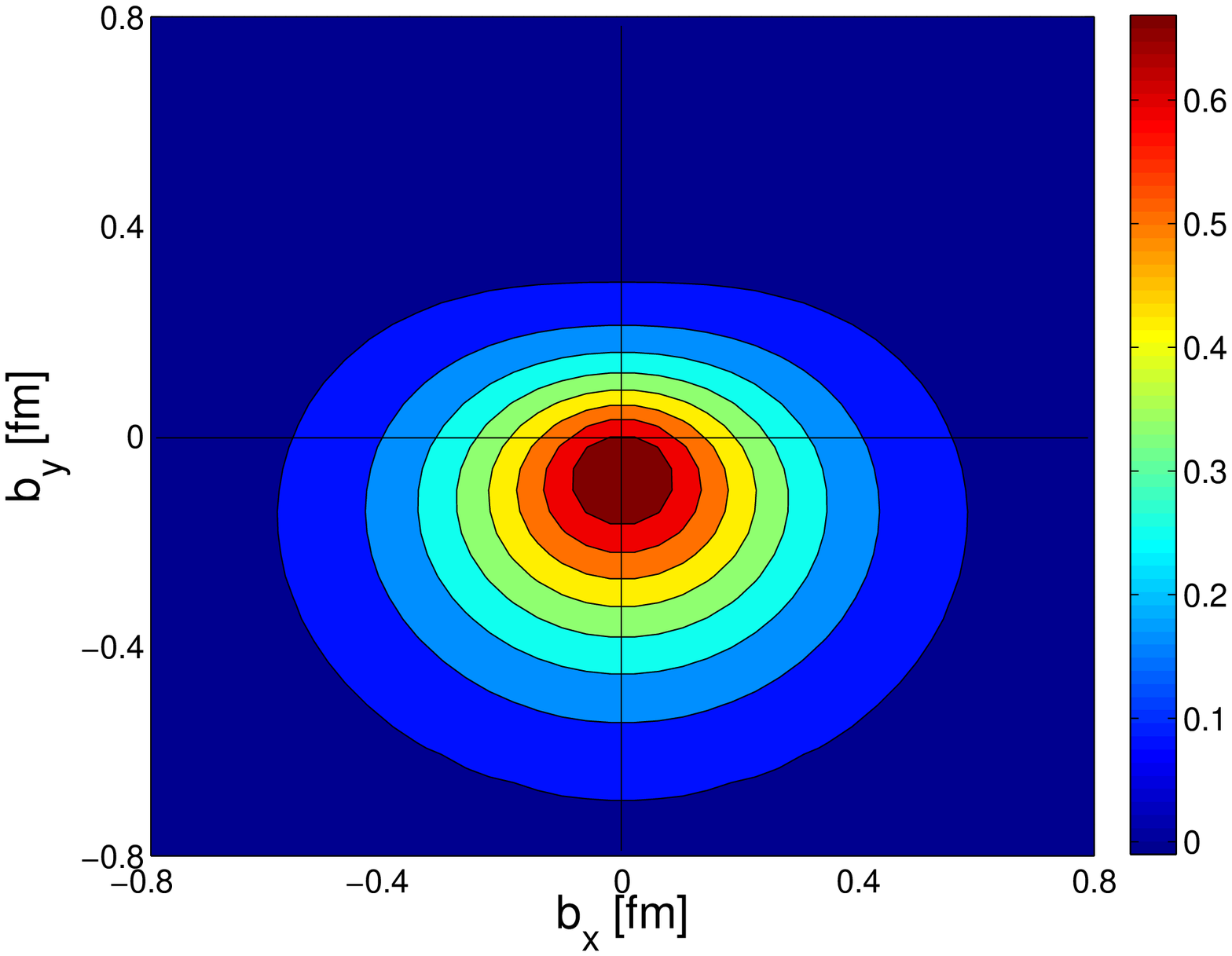}
\end{minipage}
\caption{\label{den_dI}(Color online) The longitudinal momentum densities for $d$ quark in the transverse plane for AdS/QCD Model I, upper panel  for unpolarized proton, lower panel for proton polarized along $x$-direction. (b) and (d) are the top view of (a) and (c) respectively.}
\end{figure*} 
\begin{figure*}[htbp]
\begin{minipage}[c]{0.98\textwidth}
\small{(a)}
\includegraphics[width=7cm,height=5.5cm,clip]{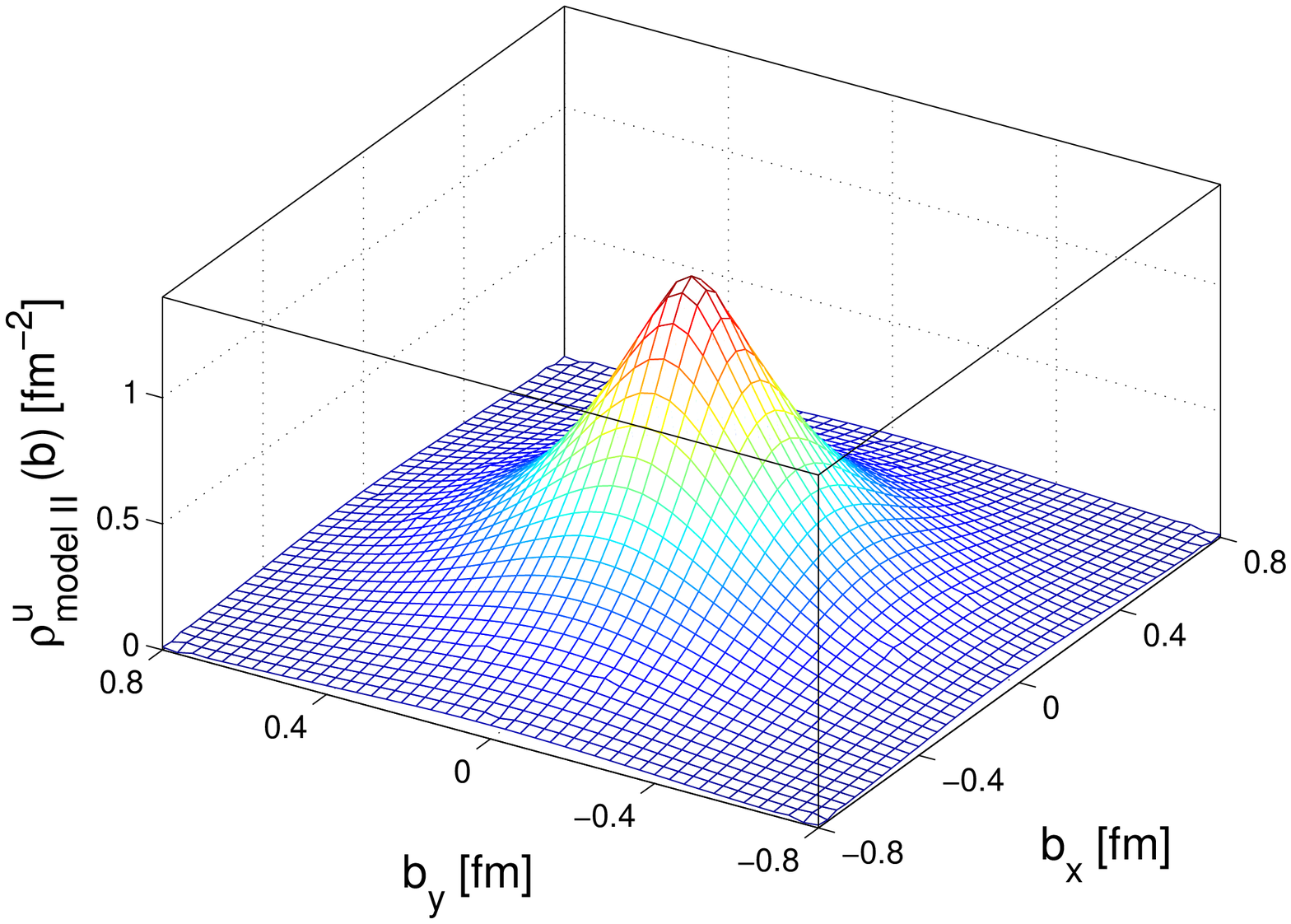}
\hspace{0.1cm}%
\small{(b)}\includegraphics[width=7cm,height=5.5cm,clip]{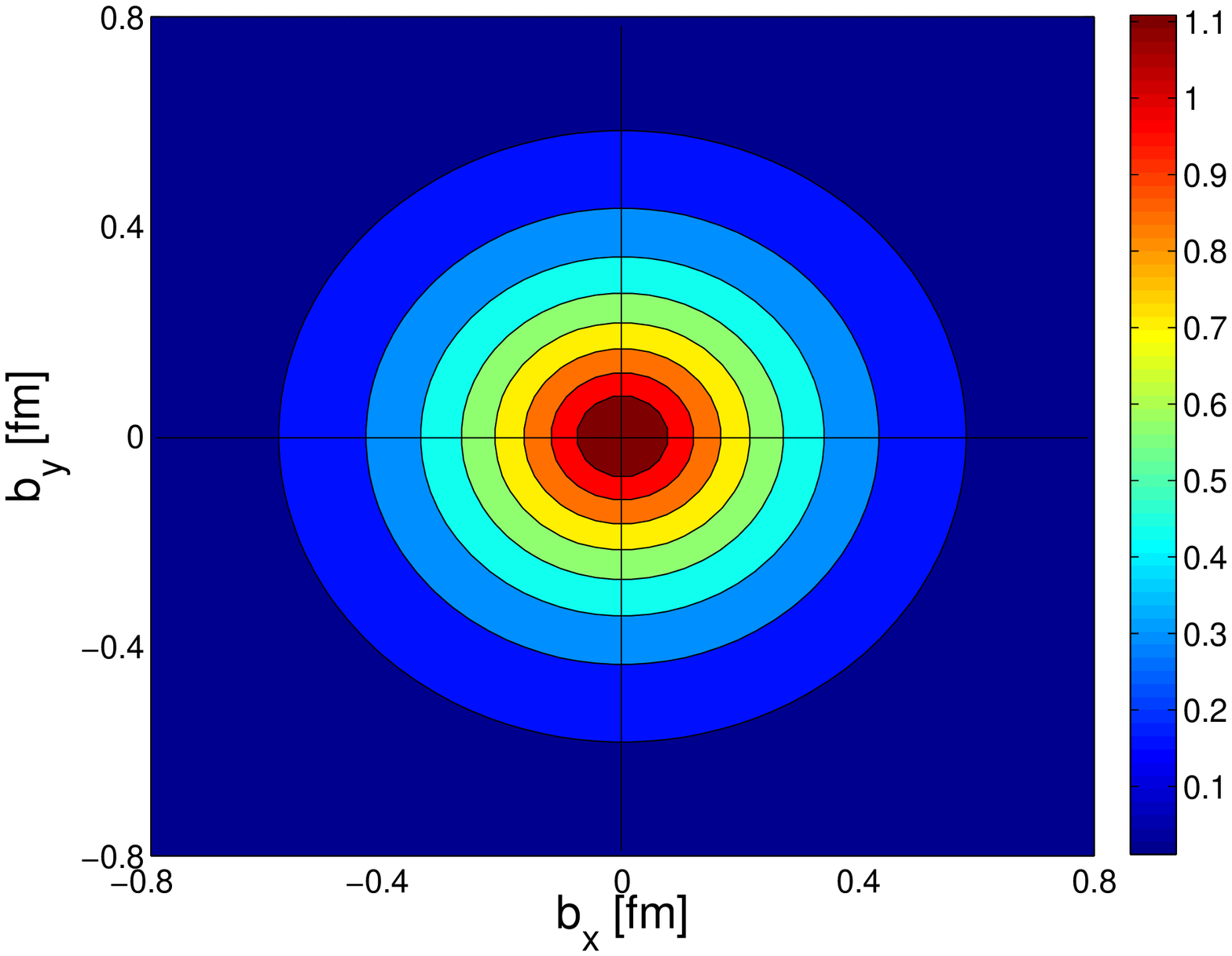}
\end{minipage}
\begin{minipage}[c]{0.98\textwidth}
\small{(c)}
\includegraphics[width=7cm,height=5.5cm,clip]{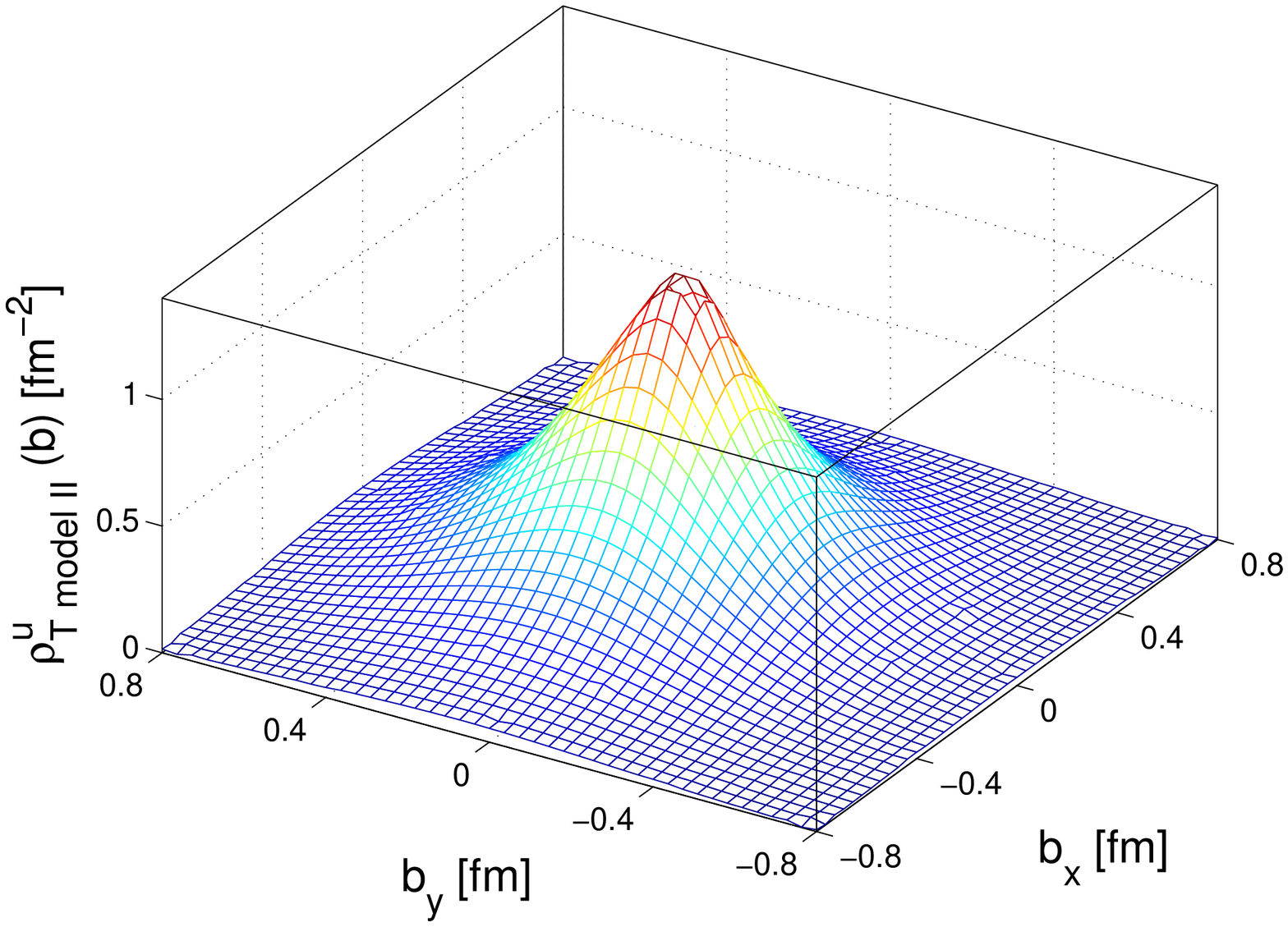}
\hspace{0.1cm}%
\small{(d)}\includegraphics[width=7cm,height=5.5cm,clip]{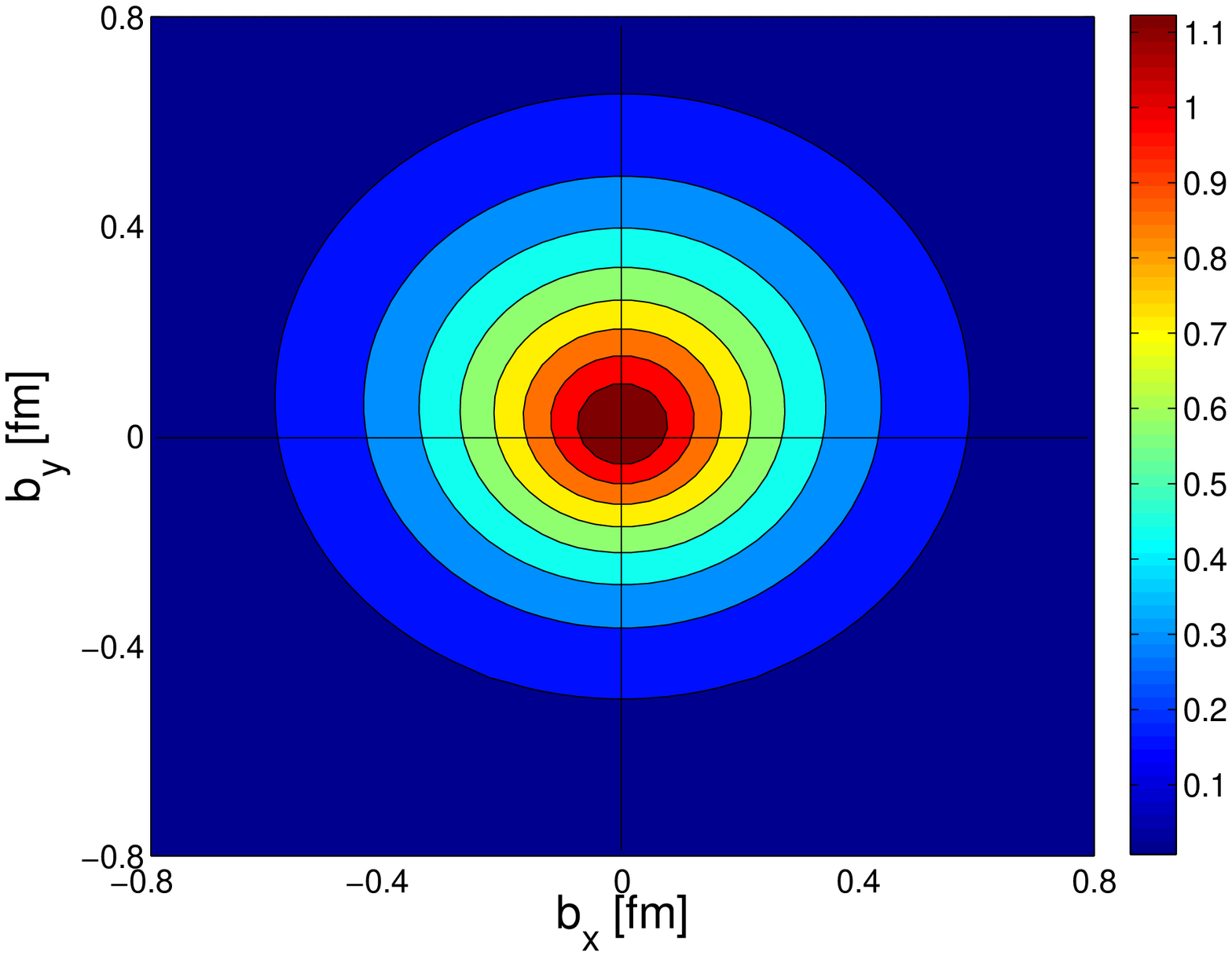}
\end{minipage}
\caption{\label{den_uII}(Color online) The longitudinal momentum densities for $u$ quark in the transverse plane for AdS/QCD Model II, upper panel  for unpolarized proton, lower panel for proton polarized along $x$-direction. (b) and (d) are the top view of (a) and (c) respectively.}
\end{figure*} 
\begin{figure*}[htbp]
\begin{minipage}[c]{0.98\textwidth}
\small{(a)}
\includegraphics[width=7cm,height=5.5cm,clip]{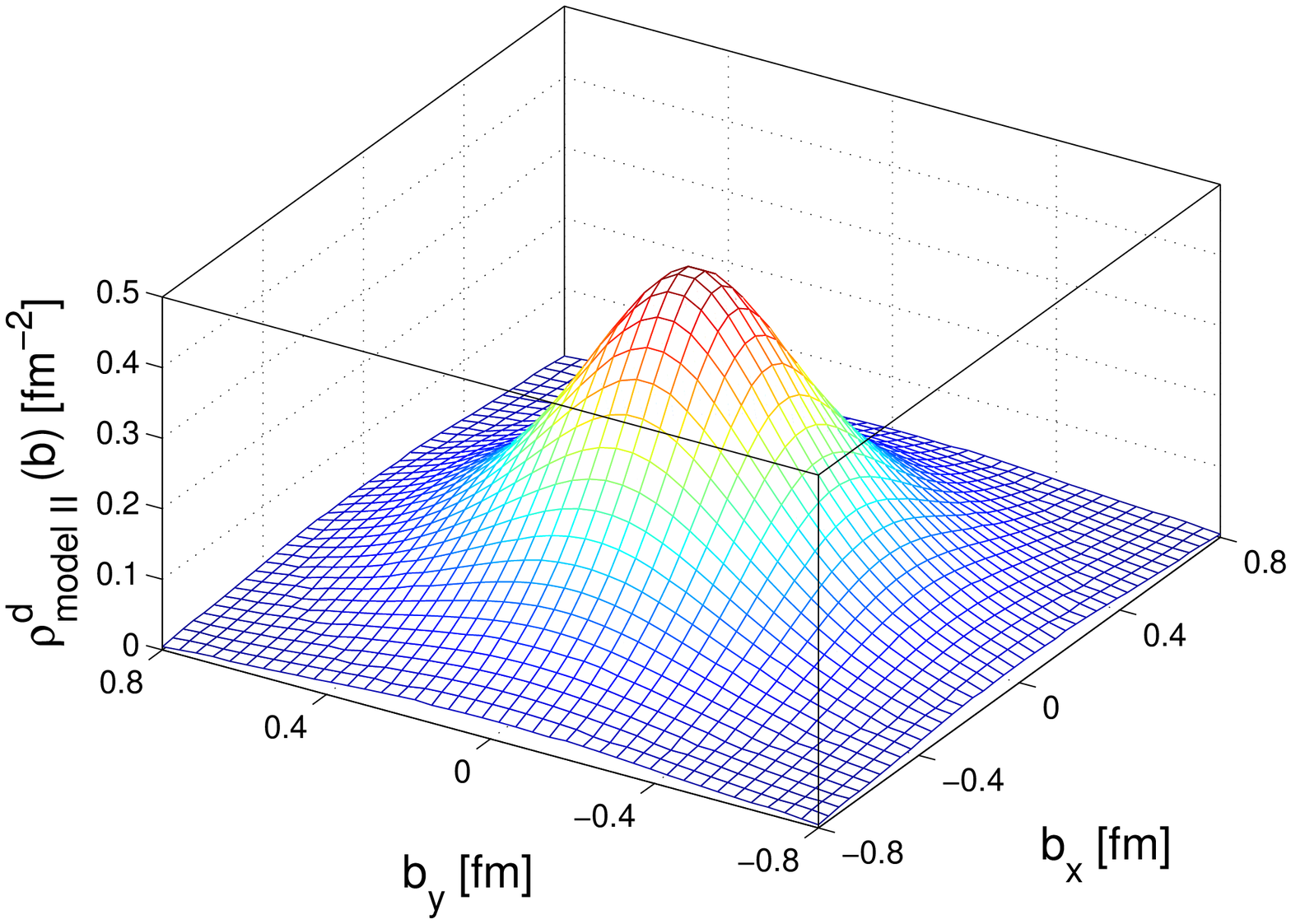}
\hspace{0.1cm}%
\small{(b)}\includegraphics[width=7cm,height=5.5cm,clip]{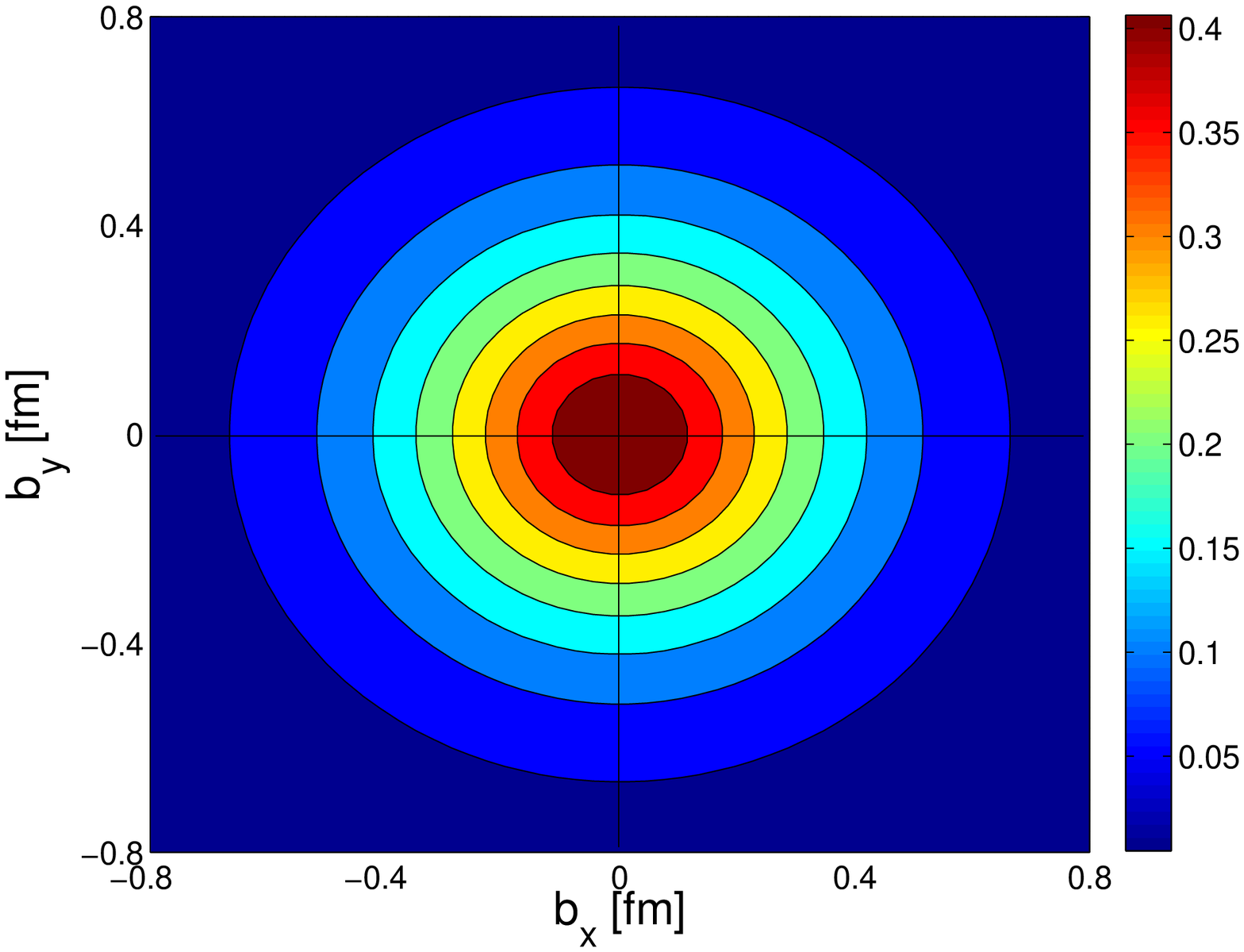}
\end{minipage}
\begin{minipage}[c]{0.98\textwidth}
\small{(c)}
\includegraphics[width=7cm,height=5.5cm,clip]{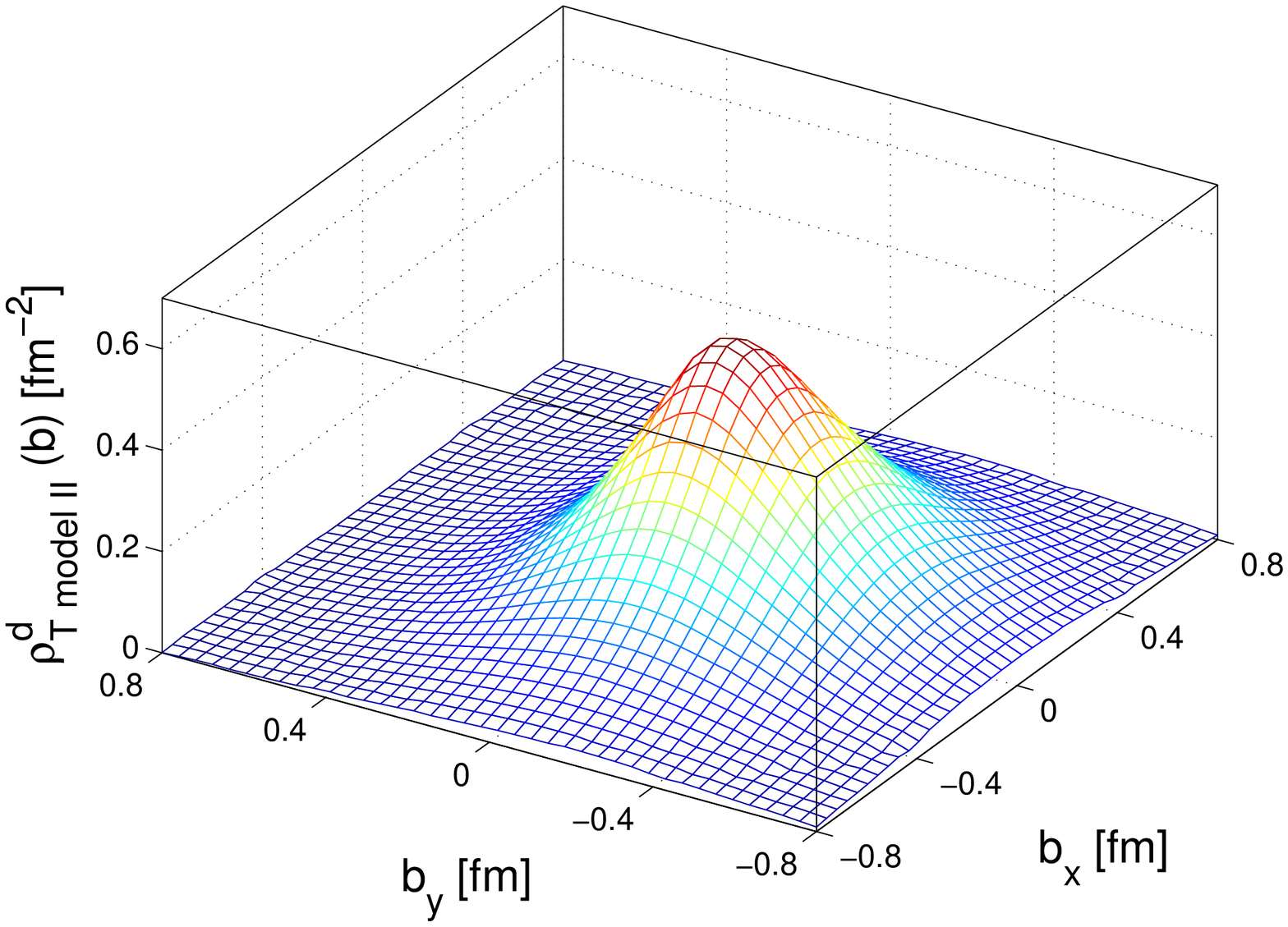}
\hspace{0.1cm}%
\small{(d)}\includegraphics[width=7cm,height=5.5cm,clip]{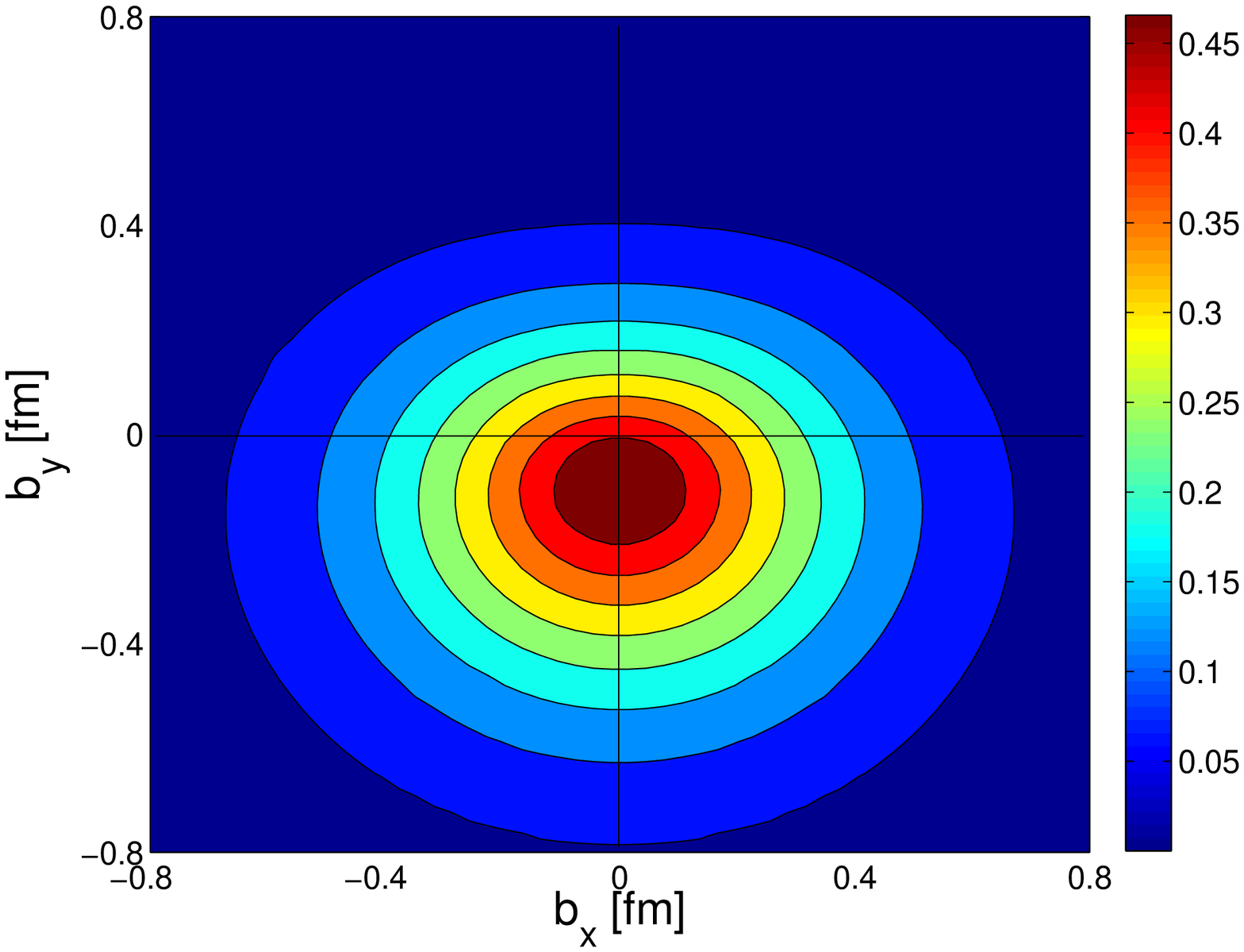}
\end{minipage}
\caption{\label{den_dII}(Color online) The longitudinal momentum densities for $d$ quark in the transverse plane for AdS/QCD Model II, upper panel  for unpolarized proton, lower panel for proton polarized along $x$-direction. (b) and (d) are the top view of (a) and (c) respectively.}
\end{figure*} 
\begin{figure*}[htbp]
\begin{minipage}[c]{0.98\textwidth}
\small{(a)}
\includegraphics[width=7.3cm,height=5.7cm,clip]{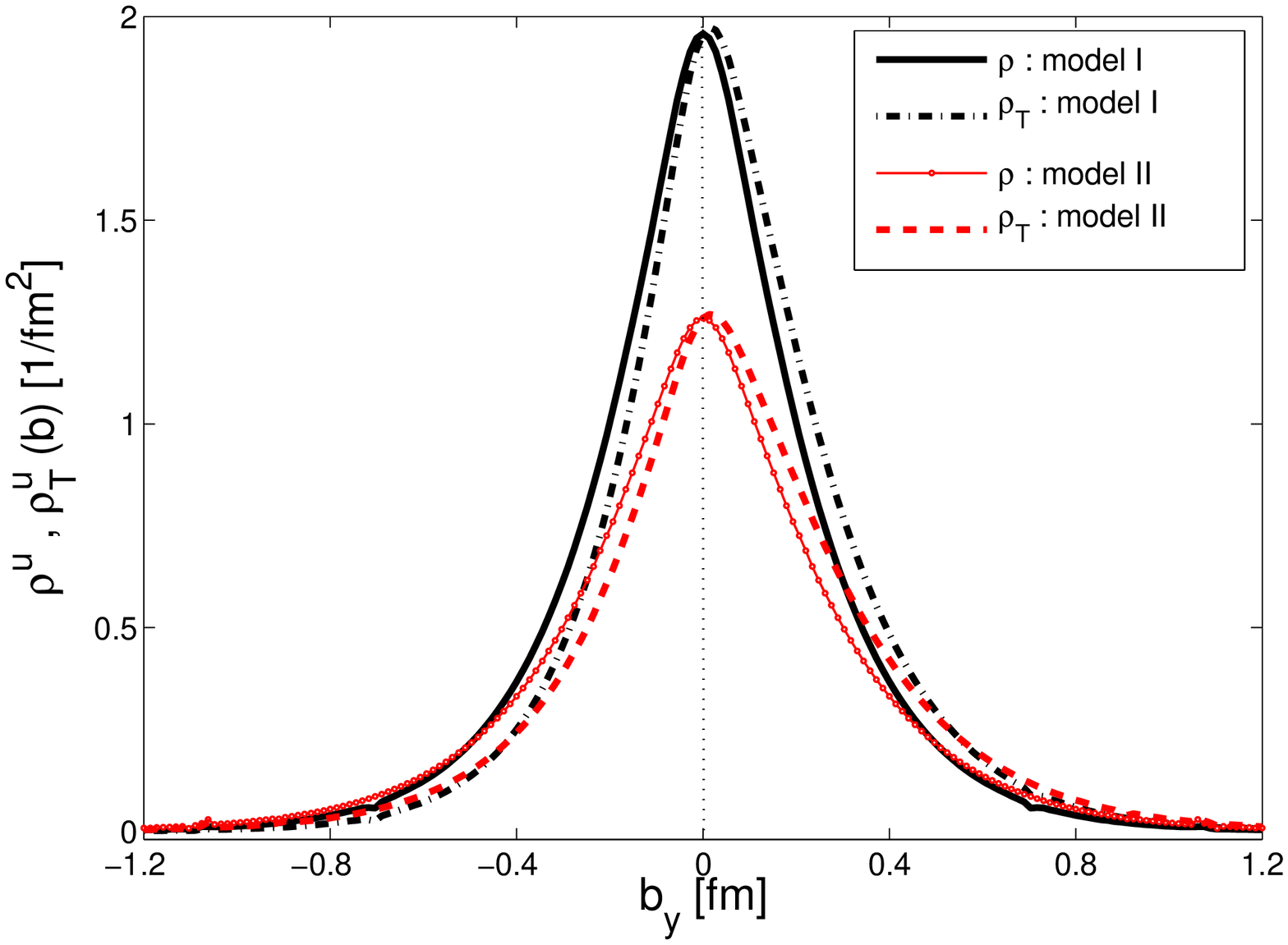}
\hspace{0.1cm}%
\small{(b)}\includegraphics[width=7.3cm,height=5.7cm,clip]{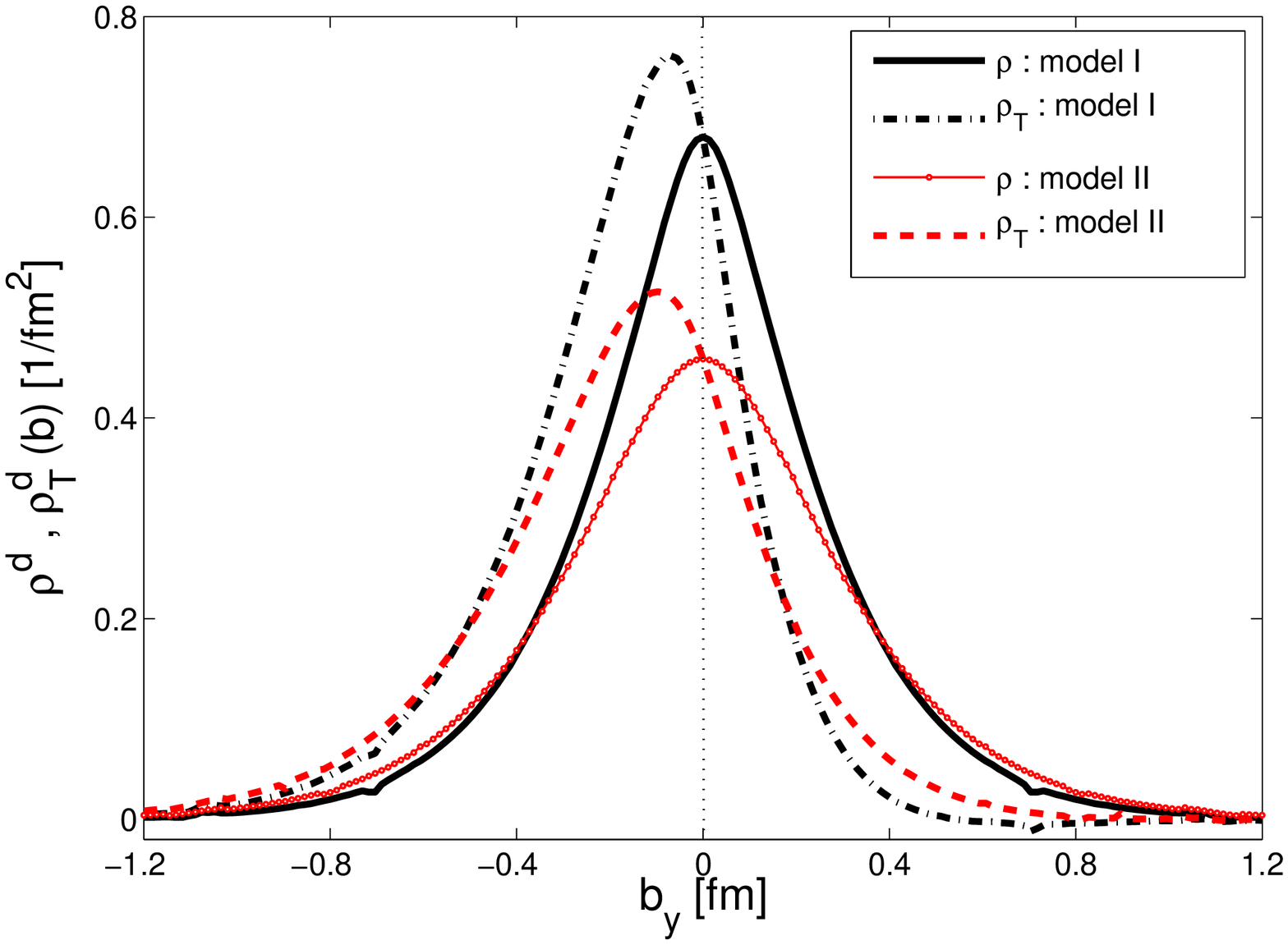}
\end{minipage}
\caption{\label{den_com}(Color online)Comparison of the longitudinal momentum densities $\rho^q$ and $\rho_T^q$ in the transverse plane between the two different AdS/QCD models, (a) for $u$ quark and (b) for $d$ quark.}
\end{figure*} 
\begin{figure*}[htbp]
\begin{minipage}[c]{0.98\textwidth}
\small{(a)}
\includegraphics[width=7cm,height=5.5cm,clip]{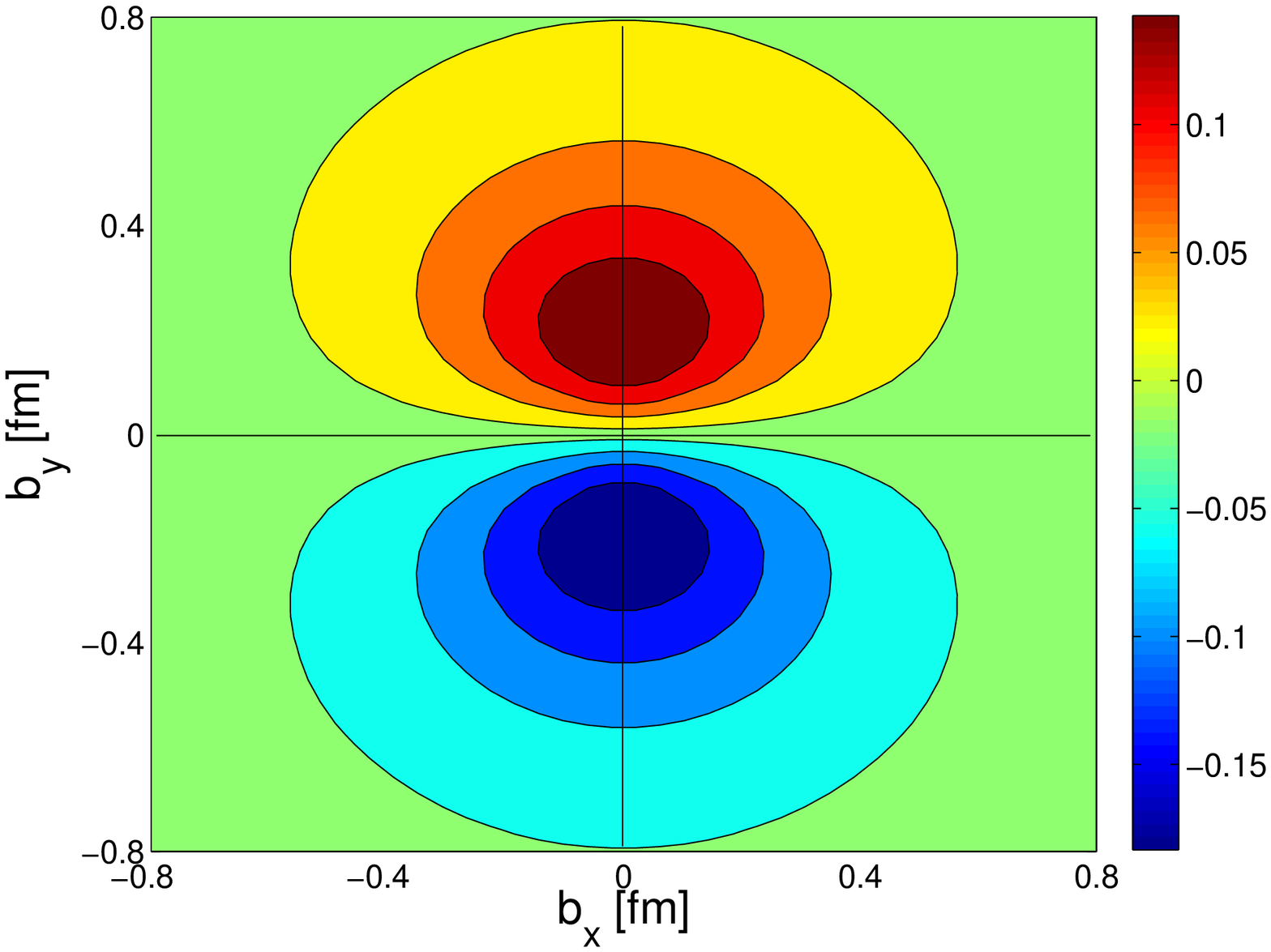}
\hspace{0.1cm}%
\small{(b)}\includegraphics[width=7cm,height=5.5cm,clip]{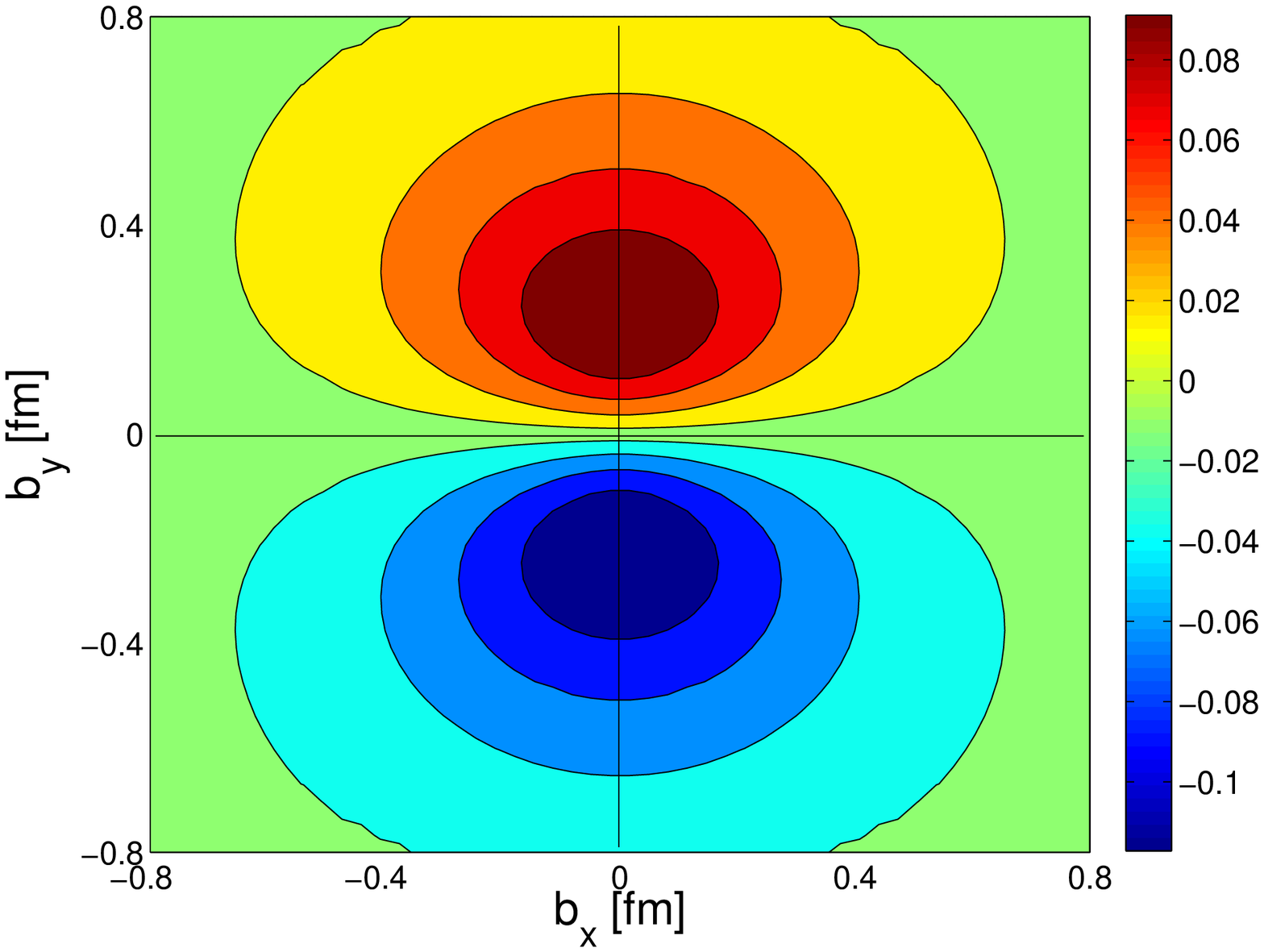}
\end{minipage}
\begin{minipage}[c]{0.98\textwidth}
\small{(c)}
\includegraphics[width=7cm,height=5.5cm,clip]{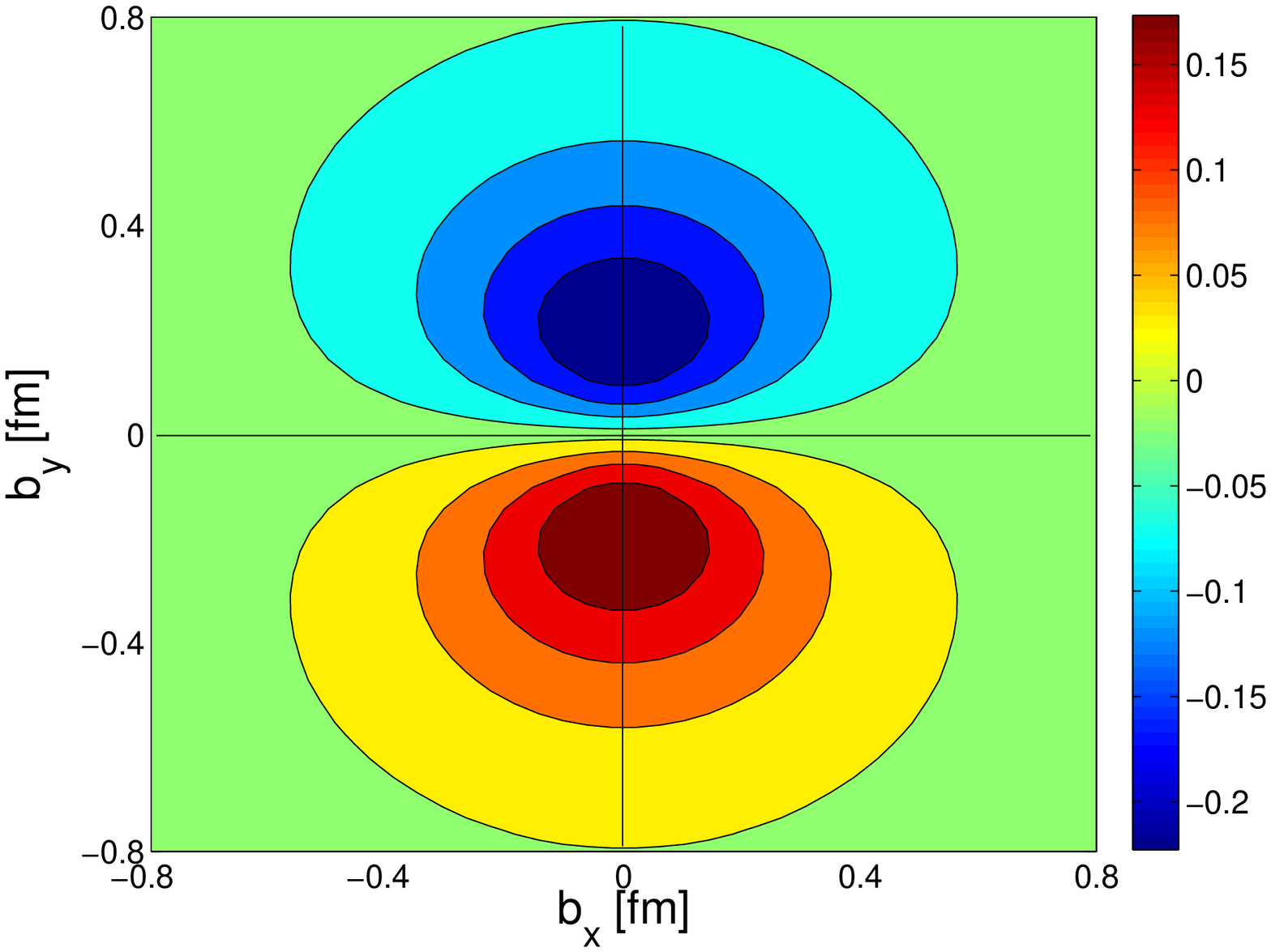}
\hspace{0.1cm}%
\small{(d)}\includegraphics[width=7cm,height=5.5cm,clip]{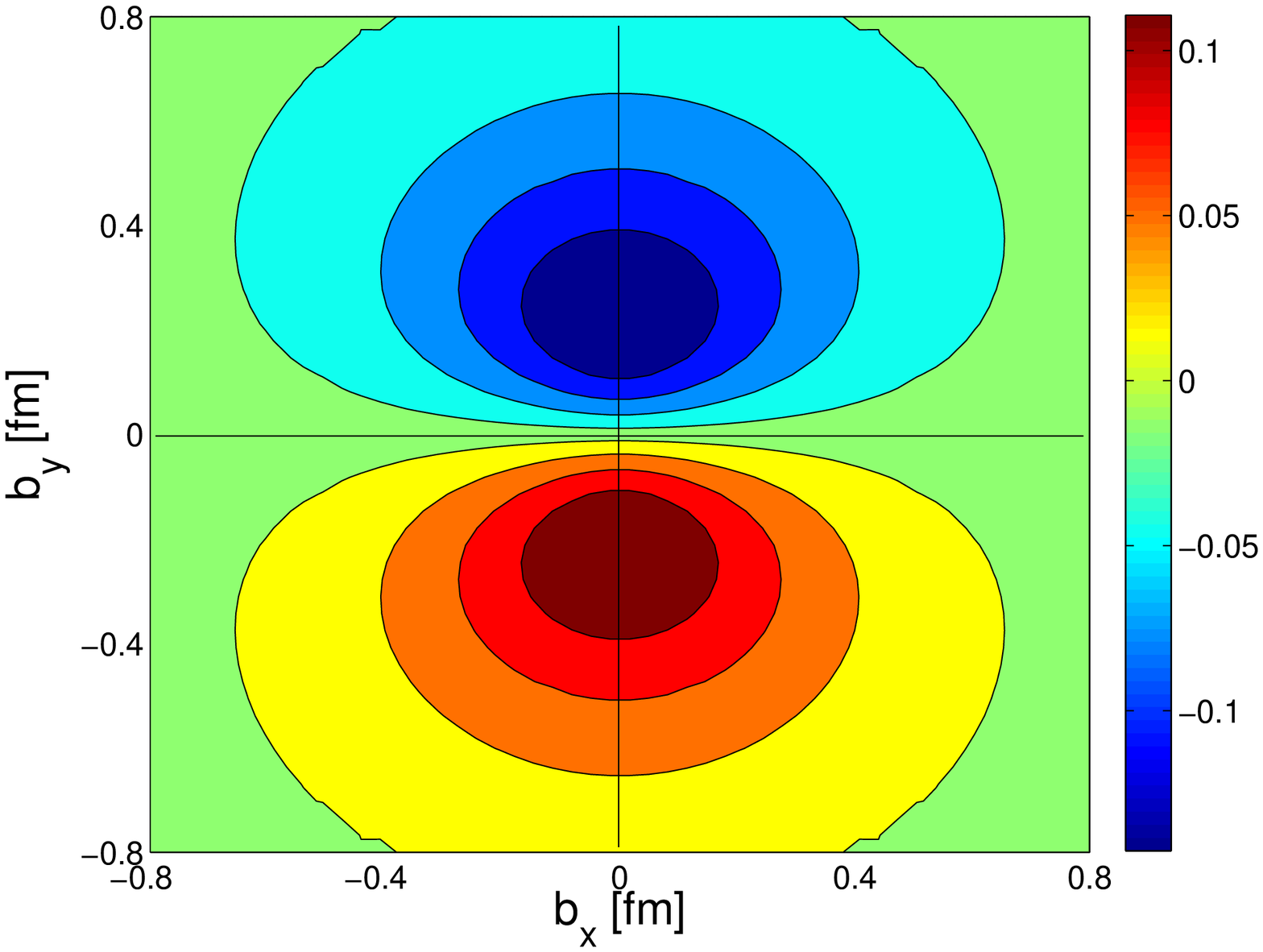}
\end{minipage}
\caption{\label{den_angular}(Color online) The momentum density asymmetry $(\rho_T(b)-\rho(b))$ in the transverse plane for  a proton polarized in $x$-direction, (a) for $u$ quark in AdS/QCD Model I (b) for $u$ quark in AdS/QCD Model II, and (c) for $d$ quark in AdS/QCD Model I (d) for $d$ quark in AdS/QCD Model II.}
\end{figure*} 

\subsection{Model II}
 The other model of the nucleon form factors was formulated by Abidin and Carlson\cite{AC}.
  A precise mapping for the spin-flip nucleon form factor using the action in Eq.(\ref{action}) is not possible. To study the Pauli form factors using holographic methods, a non-minimal electromagnetic coupling with the `anomalous' gauge invariant term has been introduced by Abidin and Carlson \cite{AC} which produces the Pauli form factors  
\be\label{F2AdS}
\frac{i}{2}~\eta_{S,V}\int d^4xdz\sqrt{g}~e^{-\Phi}  \bar\Psi
 \,  e^M_A\,  e^N_B \left[\Gamma^A, \Gamma^B\right] F_{M N}^{(S,V)}\Psi,
 \ee
 where $F_{MN}=\partial_MV_N-\partial_NV_M$ and $V_M$ is the vector field dual to electromagnetic field and $\eta_{S,V}$ are the couplings constrained by the anomalous magnetic moment of the nucleon, $\eta_p=(\eta_S+\eta_V)/2$ and $\eta_n=(\eta_S-\eta_V)/2$. The indices $S$, $V$ imply isocsalar and isovector contributions to the electromagnetic form factors.
This additional term in Eq.(\ref{F2AdS}) also provides an anomalous contribution to the Dirac form factor. In this model the form factors are given by\cite{AC}
 \be
 F_1^p(Q^2) &=& C_1(Q^2)+\eta_p C_2(Q^2),\label{F1pM2}\\
 F_1^n(Q^2)&=& \eta_n C_2(Q^2),\label{F1nM2}\\
 F_2^{p/n}(Q^2)&=& \eta_{p/n} C_3(Q^2),\label{F2pM2}
 \ee
 where the invariant functions $C_i(Q^2)$ are defined as
 \be
 C_1(Q^2)&=&\int dz~e^{-\Phi}\frac{V(Q^2,z)}{2z^3}(\psi_L^2(z)+\psi_R^2(z)),\\
 C_2(Q^2)&=&\int dz~ e^{-\Phi}\frac{{\partial}_z V(Q^2,z)}{2z^2}(\psi_L^2(z)-\psi_R^2(z)),\\
 C_3(Q^2)&=&\int dz~ e^{-\Phi}\frac{2m_nV(Q^2,z)}{z^2}\psi_L(z)\psi_R(z).
 \ee
 where $m_n$ is the mass of nucleon. The delation profile $\Phi=\kappa^2z^2$ and the normalizable wave functions $\psi_L(z)$ and $\psi_R(z)$ are the Kaluza-Klein modes, which are left and right-handed nucleon fields
\be
\psi_L(z)=\kappa^3z^4,~~~~ \psi_R(z)=\kappa^2z^3\sqrt{2}.\label{soft_modes}
\ee
The value of  $\kappa$  is fixed by simultaneous fit to proton and rho meson mass and the fit gives the value $\kappa=0.350 ~\rm GeV$.  The other parameters are determined from the normalization conditions of the Pauli form factor at $Q^2=0$ and are given by $\eta_p=0.224$ and $\eta_n=-0.239$ \cite{AC}. We refer the FFs given by Eqs. (\ref{F1pM2}-\ref{F2pM2})  as Model-II.
  
The Pauli form factors in these two models are identical,  the main difference  is in the Dirac form factor.  
In Model-II, there is an additional contribution to the Dirac form factor from the non-minimal coupling term.
It should be mentioned here that the Pauli form factors in  the AdS/QCD models are mainly of  phenomenological origin. 
The additional contribution from the non-minimal coupling to the Dirac form factor corresponds to higher twist  and not included in Model-I, while they are included in the Model-II. 

 The Dirac and Pauli FFs for the nucleons are related to the  valence GPDs by the sum rules \cite{diehl}
\be
F_1^p(t) &=& \int_0^1 dx(\frac{2}{3} H_v^u(x,t)-\frac{1}{3}H_v^d(x,t)), \nonumber\\
F_1^n(t) &=& \int_0^1 dx(\frac{2}{3} H_v^d(x,t)-\frac{1}{3}H_v^u(x,t)), \nonumber\\
F_2^p(t) &=& \int_0^1 dx(\frac{2}{3} E_v^u(x,t)-\frac{1}{3}E_v^d(x,t)), \label{FF}\\
F_2^n(t) &=& \int_0^1 dx(\frac{2}{3} E_v^d(x,t)-\frac{1}{3}E_v^u(x,t)). \nonumber 
\ee
 Here $x$ is the fraction of the light
cone momentum carried by the active quark and the GPDs for valence quark $q$ are  defined as $H_v^q(x,t)=H^q(x,0,t)+H^q(-x,0,t)$ and $ E_v^q(x,t)=E^q(x,0,t)+E^q(-x,0,t). $
Using the  integral form of the bulk-to-boundary propagator(Eq. \ref{V}) in the formulas for the FFs in AdS space for Model I (\ref{F1p}-\ref{F2}), we can rewrite the Dirac and Paula FFs as
\be\label{FF_int}
F_1^p(Q^2)&=&2\kappa^6\int_0^1 dx \int dz \frac{z^5}{(1-x)^2}x^{\frac{Q^2}{(4\kappa^2)}} e^{-\frac{\kappa^2 z^2}{(1-x)}},\nonumber\\
F_1^n(Q^2)&=&-\frac{\kappa^6}{3}\int_0^1 dx \int dz \frac{z^5(2-\kappa^2z^2)}{(1-x)^2}\nonumber\\
&\times& x^{\frac{Q^2}{(4\kappa^2)}} e^{-\frac{\kappa^2 z^2}{(1-x)}},\\
F_2^{p/n}(Q^2)&=&\kappa_{p/n}\kappa^8\int_0^1 dx \int dz \frac{z^7}{(1-x)^2}x^{\frac{Q^2}{(4\kappa^2)}} e^{-\frac{\kappa^2 z^2}{(1-x)}}.\nonumber
\ee
Comparing the integrands in Eqs.(\ref{FF}) and (\ref{FF_int}), one extracts the GPDs for Model I in the following forms
\be
H_v^u(x,t)=\frac{\kappa^6}{3}\int  &\frac{dz}{(1-x)^2}&x^{\frac{Q^2}{(4\kappa^2)}}\nonumber\\&\times& e^{-\frac{\kappa^2 z^2}{(1-x)}}z^5(\kappa^2z^2+10),\\
H_v^d(x,t)=\frac{2\kappa^6}{3}\int  &\frac{dz}{(1-x)^2}&x^{\frac{Q^2}{(4\kappa^2)}}\nonumber\\&\times& e^{-\frac{\kappa^2 z^2}{(1-x)}}z^5(\kappa^2z^2+1),\\
E_v^{u/d}(x,t)=\kappa^8\int  &\frac{dz}{(1-x)^2}&x^{\frac{Q^2}{(4\kappa^2)}} e^{-\frac{\kappa^2 z^2}{(1-x)}}z^7\kappa_{u/d},
\ee
where $\kappa_u=2\kappa_p+\kappa_n=1.673$ and $\kappa_d=\kappa_p+2\kappa_n=-2.033$ and $t=-Q^2$. Similarly we can also extract the GPDs for Model II and the GPDs in the Model II are given by
\be
H_v^u(x,t)&=&\kappa^6\int  \frac{dz}{(1-x)^2}x^{\frac{Q^2}{(4\kappa^2)}}e^{-\frac{\kappa^2 z^2}{(1-x)}}z^5\nonumber\\
\times\Big[(\kappa^2z^2&+&2)+\eta_u(\kappa^2z^2-2)(1-\frac{\kappa^2z^2x}{1-x})\Big],\\
H_v^d(x,t)&=&\kappa^6\int  \frac{dz}{(1-x)^2}x^{\frac{Q^2}{(4\kappa^2)}}e^{-\frac{\kappa^2 z^2}{(1-x)}}z^5\nonumber\\
\times\Big[\frac{1}{2}(\kappa^2z^2&+&2)+\eta_d(\kappa^2z^2-2)(1-\frac{\kappa^2z^2x}{1-x})\Big],\\
E_v^{u/d}(x,t)&=&2\sqrt{2}m_n\kappa^7\int  \frac{dz}{(1-x)^2}x^{\frac{Q^2}{(4\kappa^2)}}e^{-\frac{\kappa^2 z^2}{(1-x)}}z^7\eta_{u/d},\nonumber\\
\ee
where $\eta_u=2\eta_p+\eta_n=0.209$ and $\eta_d=\eta_p+2\eta_n=-0.254$. The GPDs in these two different models have been studied in both momentum and impact parameter spaces in \cite{CM,ModelII}. The valence  GPDs are related to the flavor GFFs  by the sum rule \cite{abidin08,selyugin}
\be 
\int_0^1 dx~x H_v^q(x,t)&=&A^q(t),\nonumber\\
\int_0^1 dx~x E_v^q(x,t)&=& B^q(t). \label{gpd}
\ee

 We use the formulas in Eq.(\ref{gpd}) to evaluate the flavor GFFs numerically from the GPDs. Being the sum of all flavors and gluon GFFs, one can get the GFFs for nucleon \cite{abidin08}. In this work, we consider only the valence quarks contributions to the nucleon. In Fig.\ref{gffs} we show the  GFFs $A(Q^2)$ and $B(Q^2)$ for $u$ and $d$ quarks. The results of AdS/QCD models are compared with a phenomenological model \cite{selyugin}.  The authors in Ref.\cite{selyugin} used the GPDs of modified Regge model \cite{guidal} by slightly changing the $t$ dependence of GPDs in the form
 \be
 H^q(x,t)&=&q(x)\exp\Big[a_+\frac{(1-x)^2}{x^{0.4}}t\Big],\\
 E^q(x,t)&=&\mathcal{E}^q(x)\exp\Big[a_-\frac{(1-x)^2}{x^{0.4}}t\Big],
 \ee
 with $\mathcal{E}^q(x)=\frac{p_q}{N_q}(1-x)^{c_q}q(x)$. The distributions $q(x)$ for $u$ and $d$ quark were taken from the MRST2002 \cite{MRST2002} global fit and all the parameters $a_{\pm}$, $p_q$, $N_q$ and $c_q$ were fixed by fitting the nucleon electromagnetic FFs with the experimental data \cite{selyugin}.
 It should be mentioned here that the flavors decompositions of nucleon electromagnetic FFs for the Model I agree well with experimental data \cite{CM2} whereas a comparative study of flavor electromagnetic FFs between these two AdS/QCD models shows that Model I is better in agreement with the experimental data than the Model II \cite{chandan_few}. In Table \ref{T}, we list the values of the GFFs at $Q^2=0$ for the two AdS/QCD models. 
 It should be noted that the values of GFFs for the zero momentum transfer in Model I are almost equal to the Model II.
 Here, the value for $(B_u(0)+B_d(0))$ is around $-0.08$ and $(A_u(0)+A_d(0))$ is around $0.9$. 
 This is because of the GPDs, we used to calculate the GFFs are valence GPDs and also there is no contribution from gluon.
When summed over all the constituents we should have $A(0)=A_q(0)+A_g(0)=1$ and $B(0)=B_q(0)+B_g(0)=0$ for hadron \cite{AC,AC4,AC5,BTgrav,BTgrav2}.  

\begin{table}[ht]
\centering 
\begin{tabular}{c c c} 
\hline 
~GFFs~~~&~~~~~~~~~Model I~~~~~~~~~& ~~~Model II~  \\ [0.5ex] 
\hline 
$A^u(0)$   & 0.6389 & 0.5868 \\
$A^d(0)$   & 0.2778 & 0.2874 \\
\hline
$B^u(0)$ & 0.4182 & 0.4180 \\ 
$B^d(0)$ & -0.5082 & -0.5079 \\
\hline 
\end{tabular} 
\caption{Gravitational FFs at $Q^2=0$} 
\label{T}
\end{table} 
\section{Longitudinal momentum densities}\label{long_mom_den}
According to the standard interpretation \cite{CM3,selyugin,miller07,vande,weiss,CM4}, in the light-cone frame with $q^+=q^1+q^3=0$, the charge and anomalous magnetization densities in the transverse plane can be interpreted with the two-dimensional Fourier transform(FT) of the Dirac and Pauli form factors. Similar to the electromagnetic densities, one can identify the gravitomagnetic density in transverse plane by taking the FT of the gravitational form factor \cite{selyugin,abidin08}.
Since the longitudinal momentum is given by the $++$ component of the energy momentum tensor
\be P^+=\int dx^-d^2x^\perp T^{++},
\ee
and the GFFs are related to the matrix element of the $++$ component of the energy momentum tensor,
it is possible to interpret the two-dimensional FT of the GFF $A(Q^2)$ as the longitudinal momentum density in the transverse plane \cite{abidin08}.
The longitudinal momentum density for a unpolarized nucleon can be defined as 
\be
\rho(b)
&=&\int \frac{d^2q_{\perp}}{(2\pi)^2}A(Q^2)e^{iq_{\perp}.b_{\perp}}\nonumber\\
&=&\int_0^\infty \frac{dQ}{2\pi}QJ_0(Qb)A(Q^2),\label{unpolarized_den}
\ee
where $b=|{b_{\perp}}|$ represents the impact parameter and $J_0$ is the cylindrical Bessel function of order zero and $Q^2=q_{\perp}^2$. The momentum density is same for both proton and neutron due to the isospin symmetry. 
The quark density gets modified by a term which involves the spin flip form factor $B(Q^2)$ when one considers a transversely polarized nucleon. The momentum density  for a transversely polarized nucleon is given by\cite{abidin08}
\be
\rho_T(b)=\rho(b)&+&\sin(\phi_b-\phi_s)\nonumber\\&\times&\int_0^\infty\frac{dQ}{2\pi}\frac{Q^2}{2M_n}J_1(bQ)B(Q^2)\label{trans_pol},
\ee
where $M_n$ is the mass of nucleon. The transverse impact parameter is denoted by $b_\perp=b(\cos\phi_b \hat{x} +\sin\phi_b\hat{y})$ and the transverse polarization of the nucleon is given by
$S_\perp=(\cos\phi_s \hat{x}+\sin\phi_s\hat{y})$. Without loss of generality, we choose the polarization of the nucleon along $x$-axis ie., $\phi_s=0$. The second term in Eq.(\ref{trans_pol}), gives the deviation from circular symmetry of the unpolarized density.

The momentum densities $\rho (b)$ for $u$ and $d$ quarks for both unpolarized and the transversely polarized nucleon in the AdS/QCD Model-I are shown in Fig.\ref{den_uI} and Fig.\ref{den_dI} respectively. Similarly for Model-II, we show the momentum densities for $u$ and $d$ quarks in Fig.\ref{den_uII} and Fig.\ref{den_dII} respectively. The unpolarized densities are axially symmetric and have the peak at the center of the nucleon$(b=0)$. For the nucleon  polarized along $x$-direction, the densities no longer have the symmetry and the peak of densities gets shifted towards positive $y$-direction for $u$ quark and opposite to $d$ quark.
For the transversely polarized nucleon, momentum densities get distorted due to the contribution coming from the second part of the Eq.(\ref{trans_pol}) which involves the gravitational FF $B(Q^2)$. Since the FF $B(Q^2)$ is positive for $u$ quark but negative for $d$ quark, the momentum densities get shifted opposite to each other for $u$ (+ve $b_y$ direction) and $d$ (-ve $b_y$ direction) and also the ratio of the contribution form $B(Q^2)$ to the momentum density with the symmetric part $\rho(b)$ is larger for $d$ quark compare to $u$ quark which causes the larger distortion for $d$ quark than $u$ quark. 
It can also be noticed that the density for $d$ quark is little wider but the height of the peak is small compare to $u$ quark in both the models. The comparison of momentum densities for the transversely polarized and unpolarized nucleon for both the models is shown in Fig.\ref{den_com}. The plots show that the shifting of the densities from the unpolarized 
symmetric densities for $d$ quark is larger than $u$ quark. Model-I gives larger momentum densities at the center of the nucleon compare to Model-II for both $u$ and $d$ quarks. Removing the axially symmetric part of the density  from $\rho_T(b)$ i.e, $(\rho_T(b)-\rho(b))$, one can find that the  angular-dependent part of the density(i.e. distortion from the symmetry)  displays a dipole pattern (Fig.\ref{den_angular}). The angular-dependent part of the density for $u$ and $d$ quarks for the Model-I are shown in Fig. \ref{den_angular}(a) and Fig. \ref{den_angular}(c). We show the same for Model-II in Fig. \ref{den_angular}(b) and Fig. \ref{den_angular}(d).  The plots show the dipole pattern but it is  broader for Model-II than Model-I. The sign of the angular-dependent part of the density for $u$  quark is  opposite to $d$ quark.

\section{Summary}\label{summary}
In this paper, we have evaluated the flavor gravitational form factor in two different soft-wall models in AdS/QCD. We have shown explicit $Q^2$ behavior of the gravitational form factors in these models and compare with a phenomenological model~\cite{selyugin}. Though both the models provide almost same values of GFFs for the zero momentum transfer $(Q^2=0)$, Model-I is better in agreement with the phenomenological model compare to Model-II. For non-zero $Q^2$, we have presented a comparative study of the longitudinal momentum density ($p^+$ density) in the transverse plane in these two models. We consider both unpolarized and transversely polarized nucleon in this work. The unpolarized densities are axially symmetric in transverse plane while for the transversely polarized nucleons they become distorted. The densities get shifted towards $y$-direction if the nucleon is polarized along $x$ direction. For transversely polarized nucleon, the asymmetries in the distributions are shown to be dipolar in nature. 
Model-I shows larger momentum density than Model-II at the center of the nucleon. The asymmetries in the distributions for Model-II is broader but less in magnitude compare to Model-I. The asymmetry in $d$ quark momentum density is found to be stronger than that for $u$ quark and shifted in opposite direction to each other.

\section{Acknowledgements}
The author thanks Dipankar Chakrabarti for critically
reading the manuscript and giving valuable suggestions and also for useful discussions.
\appendix
\section{Nucleon momentum density in AdS/QCD}\label{appa_N_den}
\begin{figure*}[htbp]
\begin{minipage}[c]{0.98\textwidth}
\small{(a)}
\includegraphics[width=7cm,height=5.5cm,clip]{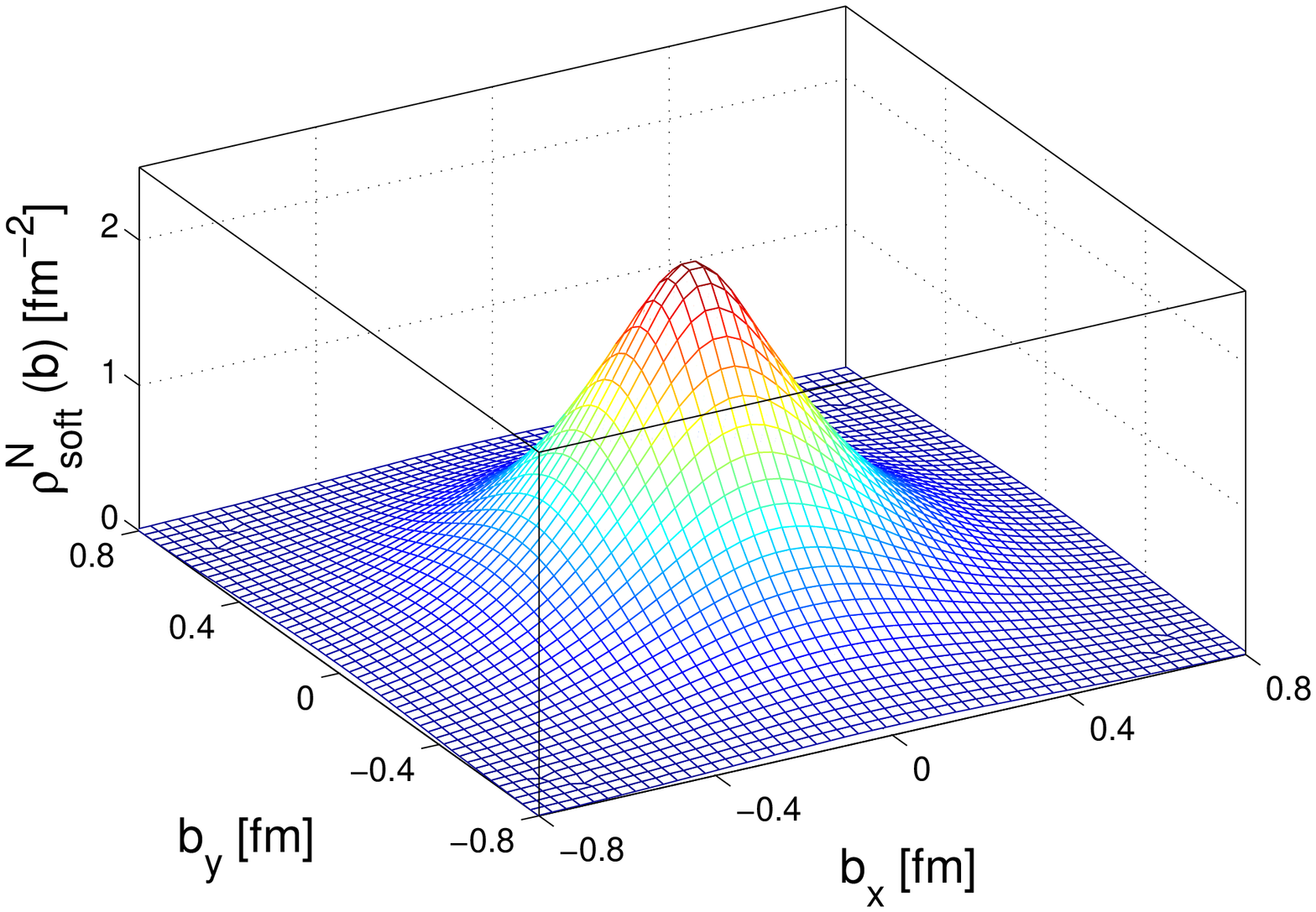}
\hspace{0.1cm}%
\small{(b)}\includegraphics[width=7cm,height=5.5cm,clip]{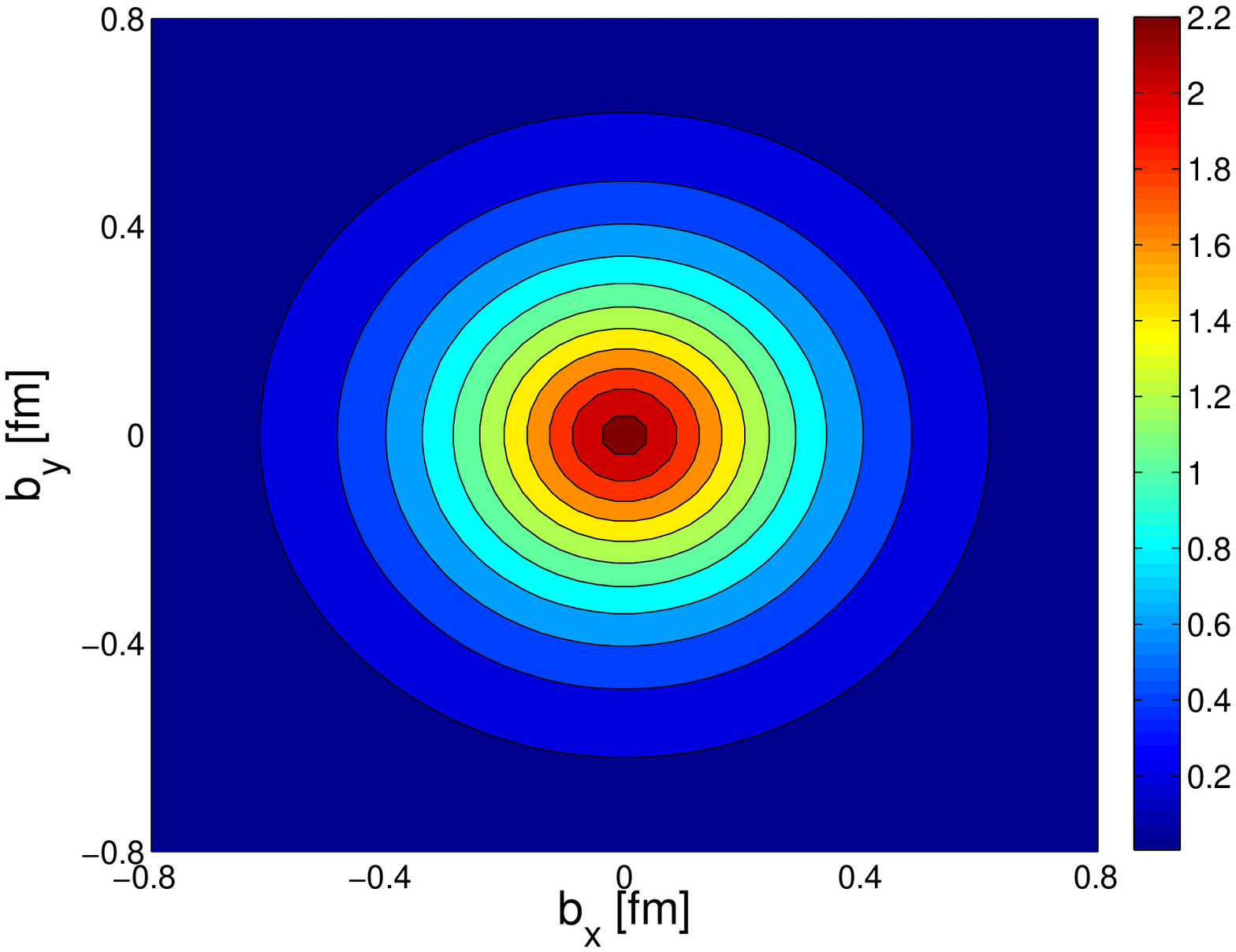}
\end{minipage}
\begin{minipage}[c]{0.98\textwidth}
\small{(c)}
\includegraphics[width=7cm,height=5.5cm,clip]{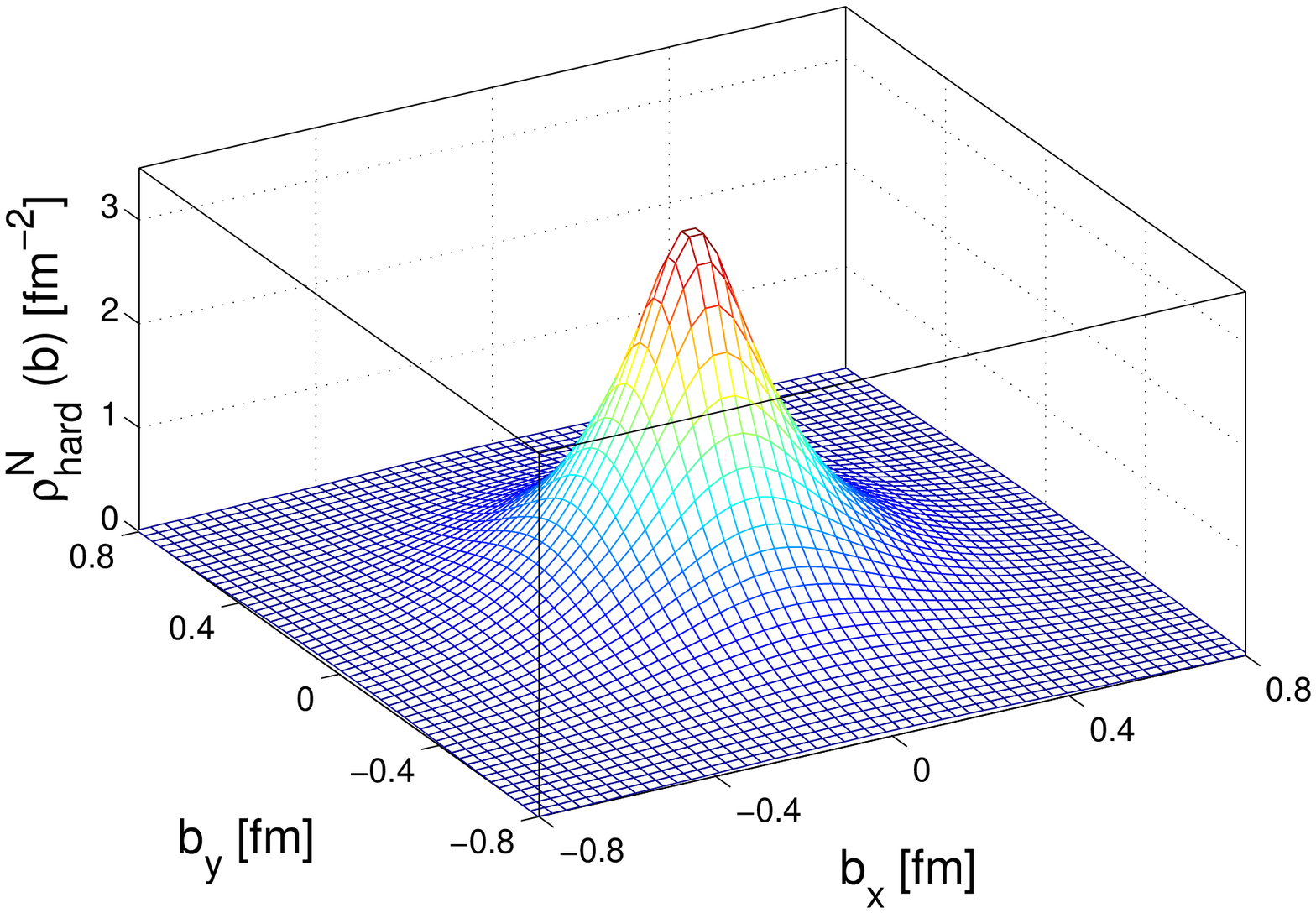}
\hspace{0.1cm}%
\small{(d)}\includegraphics[width=7cm,height=5.5cm,clip]{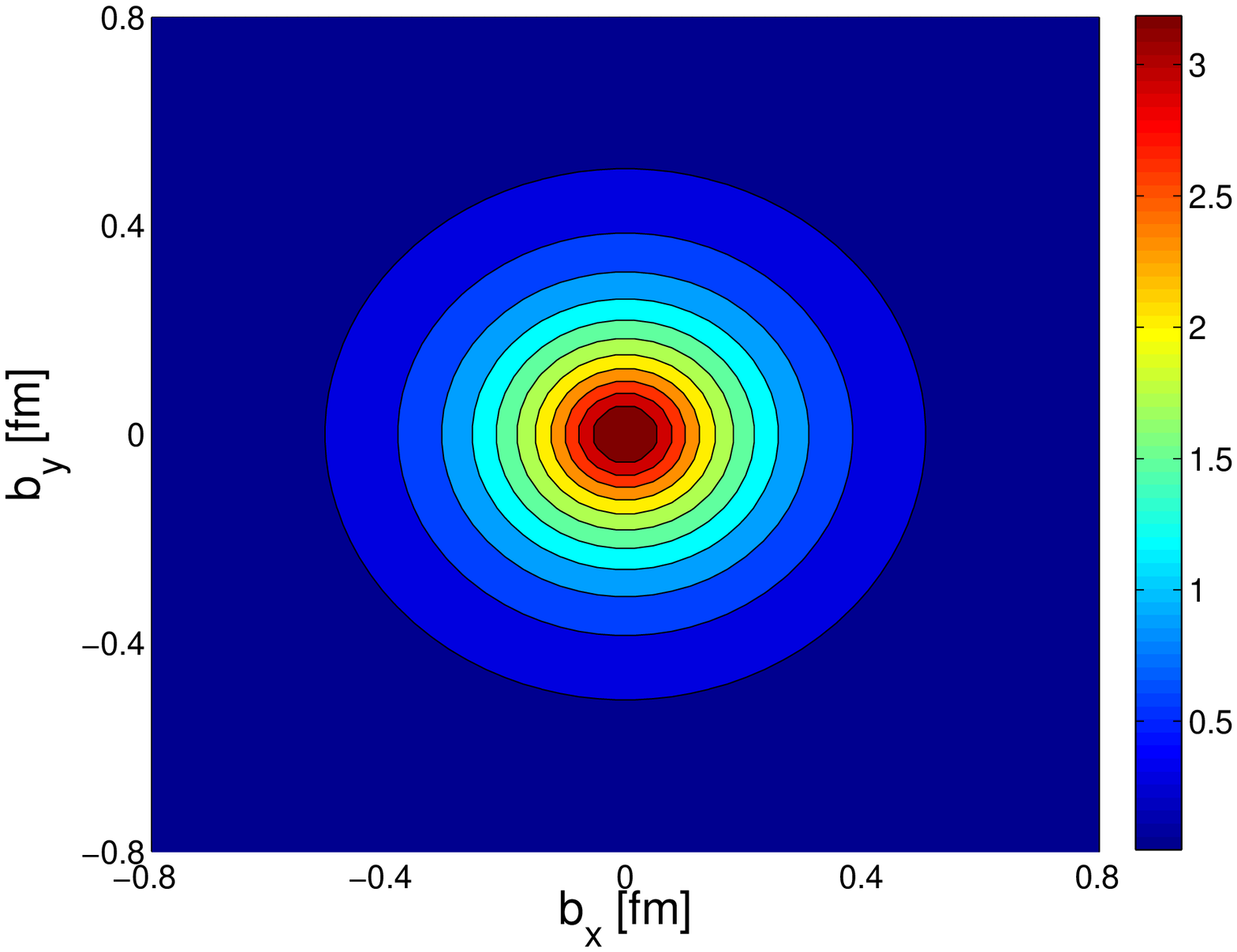}
\end{minipage}
\caption{\label{den_nucleon}(Color online) The longitudinal momentum density for nucleon in the transverse plane for, (a) soft-wall and (c) hard-wall AdS/QCD models. (b) and (d) are the top view of (a) and (c) respectively.}
\end{figure*} 
\begin{figure*}[htbp]
\includegraphics[width=8cm,height=6cm,clip]{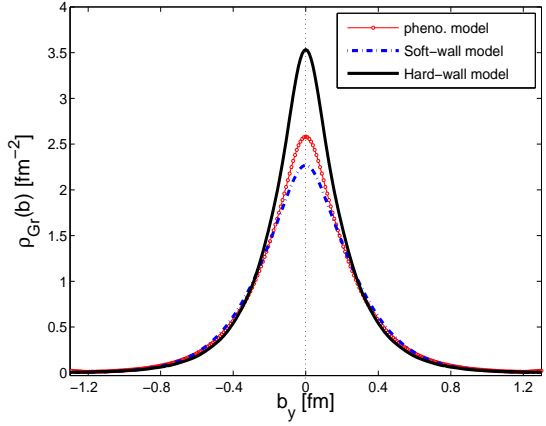}
\caption{\label{den_nucleon_com}(Color online)Comparison of the longitudinal momentum density for nucleon. The blue line with circle and the solid black lines denote the soft and hard-wall AdS/QCD models and the red dashed line represents a phenomenological medel~\cite{abidin08}.}
\end{figure*} 
To calculate the nucleon gravitational form factor, one must consider a gravity-dilation action \cite{batell08} in addition the AdS/QCD action. After perturbing the metric from its static solution according to $\eta_{\mu\nu} \to \eta_{\mu\nu} + h_{\mu\nu}$, the $5D$ gravitational action in the second order perturbation becomes \cite{AC}
\be
S_{Gr}=-\int d^5x\frac{e^{-2\kappa^2 z^2}}{4z^3}(\partial_z h_{\mu\nu}\partial^z h^{\mu\nu}+h_{\mu\nu}\square h^{\mu\nu}),
\ee
where the transverse-traceless gauge $\partial^{\mu} h_{\mu\nu}=h^{\mu}_{\mu}=0$. The profile function of the metric perturbation satisfies the following equation
\be
\Big[\partial_z\Big(\frac{e^{-2\kappa^2 z^2}}{z^3}\partial_z\Big)+\frac{e^{-2\kappa^2 z^2}}{z^3}p^2\Big]h(p,z)=0.
\ee
The solution of the profile function $H(Q,z)\equiv h(q^2=-Q^2,z)$ for the soft-wall AdS/QCD model is given by \cite{AC}
\be
H(Q,z)=a'(a'+1)\int_0^1 &dx& ~x^{a'-1}(1-x)\nonumber\\&\times&\exp\Big({-\frac{\kappa^2z^2x}{1-x}}\Big),
\ee
where $a'=\frac{Q^2}{8\kappa^2}$.
The gravitational form factor for the nucleon in AdS/QCD model has been evaluated in~\cite{AC} as
\be
A(Q^2)=\int dz\frac{e^{-\kappa^2z^2}}{2z^3}H(Q,z)(\psi_L^2(z)+\psi_R^2(z)).\label{N_GFF}
\ee
The normalizable nucleon wave-functions $\psi_L(z)$ and $\psi_R(z)$ for the soft-wall AdS/QCD model are given in Eq.(\ref{soft_modes}). The integration region in Eq.(\ref{N_GFF}) spans from $0$ to infinity. 

In the hard-wall AdS/QCD model the scale parameter $\kappa=0$ and the limit of the $z$ integration in Eq.(\ref{N_GFF}) is zero to the cutoff value $z_0=(0.245~\rm{GeV})^{-1}$. The upper cutoff was fixed in Ref.\cite{AC} to determine the nucleon and rho-meson masses. The profile function $H(Q,z)$ for the hard-wall AdS/QCD model is given by \cite{AC4}
\be
H(Q,z)=\frac{(Qz)^2}{2}\Big[\frac{K_1(Qz_0)}{I_1(Qz_0)}I_2(Qz)+K_2(Qz)\Big],
\ee
and the normalizable modes $\psi_L(z)$ and $\psi_R(z)$ in the hard-wall AdS/QCD model are \cite{AC}
\be
\psi_L(z)&=&\frac{\sqrt{2}z^2J_2(m_nz)}{z_0J_2(m_nz_0)}, \\  \psi_R(z)&=&\frac{\sqrt{2}z^2J_1(m_nz)}{z_0J_2(m_nz_0)}.
\ee
Using the GFF $A(Q^2)$ calculated in the both soft and hard-wall AdS/QCD models, we evaluate the longitudinal momentum density for nucleon as defined in Eq.(\ref{unpolarized_den}). The longitudinal momentum density $\rho^N(b)$ for nucleon for both the soft and hard-wall AdS/QCD models are shown in Fig.\ref{den_nucleon}. We compare the results of $\rho^N(b)$ in the soft and hard-wall AdS/QCD models with a phenomenological model~\cite{abidin08} in Fig.\ref{den_nucleon_com}. Our analysis shows that the soft-wall AdS/QCD model is in good agreement with the phenomenological model.



\end{document}